\documentclass[12pt,a4paper,final]{iopart}
\usepackage{cite}
 \usepackage{bm}
 \usepackage{stmaryrd,scalerel} 
\usepackage{subcaption}
\usepackage{graphicx}
\usepackage{iopams}
\usepackage{color}

\expandafter\let\csname equation*\endcsname\relax

\expandafter\let\csname endequation*\endcsname\relax

\usepackage{amsmath}
\usepackage[breaklinks=true,colorlinks=true,linkcolor=blue,urlcolor=blue,citecolor=blue]{hyperref}

\newcommand{\EQ}{\begin{equation}}
\newcommand{\EN}{\end{equation}}
\newcommand{\be}{\begin{equation}}
\newcommand{\ee}{\end{equation}}
\newcommand{\bea}{\begin{eqnarray}}
\newcommand{\eea}{\end{eqnarray}}

\newcommand{\im}{{\rm i}}

\begin{document}

\title[Fractional Brownian Gyrator]{Fractional Brownian Gyrator}	

\author{Alessio Squarcini$^{1,2,3}$, Alexandre Solon$^{4}$, Pascal Viot$^{4}$, and Gleb Oshanin$^{4}$}

\address{\em $^1$ Max Planck Institute for Intelligent Systems, Heisenbergstr. 3, D-70569, Stuttgart, Germany}

\address{\em $^2$IV. Institut f\"ur Theoretische Physik, Universit\"at Stuttgart, Pfaffenwaldring 57, D-70569 Stuttgart, Germany}

\address{\em $^3$Institut f\"ur Theoretische Physik, Universit\"at Innsbruck, Technikerstra\ss e 21A, A-6020, Innsbruck, Austria}

\address{\em $^4$Sorbonne Universit\'e, CNRS, Laboratoire de Physique Th\'eorique de la Mati\`ere Condens\'ee (UMR CNRS 7600), 4 Place Jussieu, 75252 Paris Cedex 05, France}

\begin{abstract}
When a physical system evolves in a thermal bath at a constant temperature, it arrives eventually to an equilibrium state whose properties are independent of the kinetic parameters and of the precise evolution scenario. This is generically not the case for a system driven out of equilibrium which, on the contrary, reaches a steady-state with properties that depend on the full details of the dynamics such as the driving noise and the energy dissipation. How the steady state depends on such parameters is in general a non-trivial question. Here, we approach this broad problem using a minimal model of a two-dimensional nano-machine, the Brownian gyrator, that consists of a trapped particle driven by fractional Gaussian noises -- a family of noises with long-ranged correlations in time and characterized by an anomalous diffusion exponent $\alpha$. When the noise is different in the different spatial directions, our fractional Brownian gyrator persistently rotates. Even if the noise is non-trivial, with long-ranged time correlations, thanks to its Gaussian nature we are able to characterize analytically the resulting nonequilibrium steady state by computing the probability density function, the probability current, its curl and the angular velocity and complement our study by numerical results.
\end{abstract}


\maketitle



\section{Introduction}
\label{introduction}
The Brownian gyrator model, that consists of a pair of coupled,
standardly defined Ornstein-Uhlenbeck processes each evolving at its
own temperature, is one of the simplest models exhibiting a
non-trivial out-of-equilibrium dynamics and, for this reason, has
received much interest in the last two decades.  It was originally
introduced as a solvable model of dynamics with two
temperatures~\cite{Exartier1999}, and it was realized only years later
that a systematic torque is generated on the
object~\cite{Filliger2007}, so that it can be seen as a minimal
nano-machine that, on average, undergoes steady gyration around the
origin. Following this remarkable observation, various facets of the
model have been studied in different contexts. In addition to the
theoretical calculation of the evolution of some relevant properties
(see {\it e.g.}
Refs.~\cite{Dotsenko2013,Fogedby2017,Bae2021,Chang2021,Alberici2021}),
several more general questions have also been addressed including the
role of cross-correlations between different degrees of freedom
coupled to different thermostats on the production of entropy
\cite{Crisanti2012}, the directionality of interactions between
cellular processes~\cite{Lahiri2017}, statistical properties of the
energy exchanged between two heat baths in electric circuits
\cite{Ciliberto2013,1742-5468-2013-12-P12014} and their spectral
properties \cite{Cerasoli2022}, dynamics in suspensions of two types
of particles exposed to different heat baths \cite{Grosberg2015} and
in laser-cooled atomic gases \cite{Mancois2018}, as well as the
synchronization of out-of-equilibrium Kuramoto oscillators
\cite{Dotsenko2019}. Furthermore, exerting external forces on a
Brownian gyrator has lead to derive a special fluctuation theorem and
to identify associated effective temperatures
\cite{Cerasoli2018,Cerasoli2022,Tyagi2020}. The latter setting with
external forces was found to be mathematically equivalent to the
bead-spring model studied via Brownian dynamics simulations in
Ref.~\cite{Battle2016} and analytically in
Refs.~\cite{Li2019,Sou2019}. Finally, different versions of a Brownian
gyrator were studied experimentally, especially from the point of view
of fluctuation
relations~\cite{Ciliberto2013,1742-5468-2013-12-P12014,argun2017experimental}.

All the above-mentioned analyses concentrated exclusively on
situations in which the noise acting on both components of the gyrator
is thermal and delta-correlated in time. However, nano-machines can be
encountered or engineered in a variety of environments including active
suspensions~\cite{di2010bacterial,zakine2017stochastic} or complex
biological mediums~\cite{needleman2017active} where the noise is
athermal and correlated in time. It is thus important to understand
the effects of correlated noise on the Brownian gyrator. A first step
was made recently in Ref.~\cite{Nascimento2021} which considered
persistent noise that is exponentially correlated in time and the
associated memory kernel for the damping so that, individually, the
noise on each component remains thermal. Even for such short-ranged
correlations, essential departures from the standard white noise model
emerge which raise the question of the effect of other types of noise,
athermal and/or with long-ranged time correlations.  From a broader
perspective, because of its simplicity, the Brownian gyrator model is
an ideal playground to understand the effects of different types of
noise correlations on the properties of a nonequilibrium steady state.
 
In the present paper, we tackle the question formulated above by
generalizing the Brownian gyrator model to a fractional Brownian
gyrator (FBG) in which the noise acting on each component is the
so-called fractional Gaussian noise (FGN)~\cite{Mandelbrot1968}. The
latter, in fact, represent a family of colored noises parametrized by
an anomalous diffusion exponent $\alpha$ such that an overdamped
particle in free space subject to a FGN performs the standard
fractional Brownian motion
\cite{Mandelbrot1968,Guggenberger2021}. When $\alpha < 1$ the
increments are anti-correlated and the dynamics is sub-diffusive
whereas for $\alpha > 1$ the increments are positively correlated
which entails a super-diffusive motion. The standard Brownian gyrator
model is recovered in the borderline case when $\alpha = 1$. Our
analysis here covers all three situations within a unified framework,
so that we can inquire for each type of correlations if they are
beneficial or, on the contrary, detrimental to rotation. In any case,
rotation can appear only if the symmetry between the two spatial
components is broken. Here, to assess independently the different
features of the noise, we discuss especially two situations: In the
first case, both noises have the same $\alpha$ but different
amplitudes while in the second case the amplitudes are equal while the
values of $\alpha$ differ. We calculate exactly the probability
density function $P(x,y;t)$ and present its explicit form in the limit
$t \to \infty$. We further compute the probability current, its curl,
and eventually calculate the ensemble-averaged angular velocity. For
each quantity, we compare our results to numerical simulations.

The paper is outlined as follows: We first introduce the dynamics and
detail the two variants of the model that breaks the symmetry using
different parameters in Sec.~\ref{Model}. In Sec.~\ref{Position} we
compute analytically the first moments of the position before giving
the full probability density function and considering numerical
applications.  We further compute some important quantities
characterizing the nonequilibrium steady state in Sec.~\ref{Rotation}:
the probability current, its curl and the mean angular
velocity. Finally, we summarize our findings in
Sec.~\ref{conc}. Details of intermediate calculations are relegated to
Appendices.

\section{The Model} \label{Model}
\subsection{Definition}
Let us first present the standard Brownian gyrator (BG) as introduced
in Ref.~\cite{Exartier1999}. We denote by $x(t)$ and $y(t)$ the
Cartesian coordinates of an overdamped particle evolving in a
potential $U(x,y)$ according to the Langevin equation
\begin{equation}
\begin{aligned}
\label{001}
\gamma\dot{x}(t) & = - \partial_{x}U(x,y) + \xi_{x}(t)  \\
\gamma\dot{y}(t) & = - \partial_{y}U(x,y) + \xi_{y}(t) \, ,
\end{aligned}
\end{equation}
where $\gamma$ is the damping coefficient,
$U(x,y)=(1/2)(k_x x^2+k_y y^2)-u xy$ is a generic harmonic potential
and $\xi_{x}(t)$ and $\xi_{y}(t)$ are additive Gaussian white noises
with zero mean.  Adopting appropriate space and time units, we can
choose without loss of generality $k_x=k_y=1$ and $\gamma=1$ so that
the dynamics rewrite
\begin{equation}
\begin{aligned}
\label{001a}
\dot{x}(t) & = - x(t) + u y(t) + \xi_{x}(t) \\
\dot{y}(t) & = - y(t) + u x(t) + \xi_{y}(t) \, .
\end{aligned}
\end{equation}
with noise variance
\begin{equation}
\label{001aa}
\langle \xi_{i}(t_{1}) \xi_{j}(t_{2}) \rangle = 2 T_{j} \delta_{i,j} \delta(t_{1}-t_{2}) \, , \qquad i,j =x,y \,.
\end{equation}
We are interested in the case when the particle is trapped, which
requires $|u|<1$. When $T_x\neq T_y$, the system is out of equilibrium
and exhibits a persistent rotation in steady
state~\cite{Filliger2007}.

In this paper, we generalize the BG model by considering the situation
in which $\xi_{x}(t)$ and $\xi_{y}(t)$ are two fractional Gaussian
noises (FGNs) with different parameters so that their correlation
function reads
\begin{equation}
\label{001aaa}
\langle \xi_{i}(t_{1}) \xi_{j}(t_{2}) \rangle = T_{j} \delta_{ij} N_{\alpha_{j}}(t_{1},t_{2}) \, .
\end{equation}
The kernel $N_{\alpha_{j}}(t_{1},t_{2})$ is defined such that in free
space, $\dot x_{\rm free}=\xi_x(t)$ yields the so-called fractional
Brownian for $x_{\rm free}$ (see, {\it e.g.}, \cite{Jeon2010}):
\begin{align}
 \nonumber
\langle x_{\rm free}(t_{1}) x_{\rm free}(t_{2}) \rangle & = T_x \int_{0}^{t_{1}}\textrm{d}\tau_{1} \int_{0}^{t_{2}}\textrm{d}\tau_{2} \, N_{\alpha x}(\tau_{1},\tau_{2}) \\
                                                        & = T_x ( t_{1}^{\alpha} + t_{2}^{\alpha} - |t_{1}-t_{2}|^{\alpha} ) \, .
                                                          \label{003}
\end{align}
In particular, the mean squared displacement exhibits an anomalous diffusion
\begin{equation}
\langle x_{\rm free}^{2}(t) \rangle = 2 T_x t^{\alpha} \, ,
\end{equation}
where $\alpha \in (0,2)$ is the anomalous diffusion
exponent. $\alpha=1$ corresponds to the standard Brownian motion while
$\alpha<1$ and $\alpha>1$ generate sub-diffusion and super-diffusion
respectively.  The explicit expression of the kernel $N_{\alpha_{j}}$
that satisfies Eq.~(\ref{003}) for arbitrary $\alpha \in (0,2)$ is not
needed for the calculations presented in this paper. The interested
reader can find detailed discussions on this topic in
Refs.\cite{Jeon2010,Jeon2012,Jeon2012a,Jeon2014}.

Let us note that, although in the following we will call abusively
$T_x$ and $T_y$ ``temperatures'', one should keep in mind that they
are not physical temperatures but rather only the noise
amplitudes. Indeed, even if the components are independent ($u=0$),
each component is out of equilibrium since the noise and damping do
not satisfy the fluctuation-dissipation theorem, except when
$\alpha=1$.

Numerical simulations are performed by solving Eq.~(\ref{001a}) using
an Euler–Maruyama algorithm. The FGN is generated by using the Davies
and Harte method~\cite{Davies1987} and is provided by the Stochastic
python module.

\subsection{Two variants}

Although all our analytical results can be derived for the general FBG
introduced above, it will help to discuss specifically two variants of
the model in which we vary different parameters. In the first one
(Model 1) the anomalous diffusion exponent is unique,
$\alpha_{x}=\alpha_{y} \equiv \alpha$, but the temperatures $T_{x}$
and $T_{y}$ are different, so that the noise correlations read
\begin{equation}
	\label{002}
	{\rm Model\,1:} \qquad \langle \xi_{x}(t_{1}) \xi_{x}(t_{1}) \rangle = T_{x} N_{\alpha}(t_{1},t_{2}) \, , \qquad \langle \xi_{y}(t_{1}) \xi_{y}(t_{1}) \rangle = T_{y} N_{\alpha}(t_{1},t_{2}) \, .
\end{equation}
In Model 2, the temperature $T_{x}=T_{y} \equiv T$ is the same for
both components but the anomalous diffusion exponents are different,
so that
\begin{equation}
	{\rm Model\,2:} \qquad \langle \xi_{x}(t_{1}) \xi_{x}(t_{1}) \rangle = T N_{\alpha_{x}}(t_{1},t_{2}) \, , \qquad \langle \xi_{y}(t_{1}) \xi_{y}(t_{1}) \rangle = T N_{\alpha_{y}}(t_{1},t_{2}) \, .
\end{equation}

\section{Spatial distribution}
\label{Position}

We first solve the equations of motion for a given realization of the
noise which will allow us to obtain any moment of the position
distribution. Thanks to their linear structure, the equations of
motion Eq.~(\ref{001a}) are readily integrated to give
\begin{equation}
\begin{aligned}
\label{004}
x(t) & = \int_{0}^{t}\textrm{d}\tau \, Q_{c}(t-\tau) \xi_{x}(\tau) + \int_{0}^{t}\textrm{d}\tau \, Q_{s}(t-\tau) \xi_{y}(\tau) \\
y(t) & = \int_{0}^{t}\textrm{d}\tau \, Q_{s}(t-\tau) \xi_{x}(\tau) + \int_{0}^{t}\textrm{d}\tau \, Q_{c}(t-\tau) \xi_{y}(\tau) \, ,
\end{aligned}
\end{equation}
with the kernels
\begin{equation}
\label{005}
Q_{c}(t) = \textrm{e}^{-t} \cosh(u t) \, , \qquad Q_{s}(t) = \textrm{e}^{-t} \sinh(u t) \, .
\end{equation}
For simplicity, and since we are interested only in the steady-state
properties, we chose the initial condition $x(0)=y(0)=0$. Since the
noises have zero mean it follows that
$\langle x(t) \rangle = \langle y(t) \rangle=0$ at all time.

In the following, we first compute the second moment of the position
in Sec.~\ref{Position_2nd-moment} and the full probability density
function in Sec.~\ref{Position_proba} and analyze quantitatively the
effect of the FGN on the spatial distribution in
Sec.~\ref{Position_quantitative}.

\subsection{Mean squared displacement and cross correlations}
\label{Position_2nd-moment}
The second moments of the position distribution can be computed at all
time from Eq.~(\ref{004}) and the expression of the noise correlation
Eq.~(\ref{003}). They take the form
\begin{align}
\label{016a}
\langle x^{2}(t) \rangle & = A_{x}(t) T_{x} + B_{y}(t) T_{y}  \\
\label{016b}
\langle y^{2}(t) \rangle & = B_{x}(t) T_{x} + A_{y}(t) T_{y}  \\
\label{016c}
\langle x(t) y(t) \rangle & = C_{x}(t) T_{x} + C_{y}(t) T_{y}  \, ,
\end{align}
where $A_{j}(t)$, $B_{j}(t)$, and $C_{j}(t)$ are given by the following integrals
\begin{align}
\label{011a}
A_{j}(t) & = \int_{0}^{t}\textrm{d}\tau_{1} \int_{0}^{t}\textrm{d}\tau_{2} \, Q_{c}(t-\tau_{1}) Q_{c}(t-\tau_{2}) N_{\alpha_{j}}(\tau_{1},\tau_{2}) \\
\label{011b}
B_{j}(t) & = \int_{0}^{t}\textrm{d}\tau_{1} \int_{0}^{t}\textrm{d}\tau_{2} \, Q_{s}(t-\tau_{1}) Q_{s}(t-\tau_{2}) N_{\alpha_{j}}(\tau_{1},\tau_{2}) \\ 
\label{011c}
C_{j}(t) & = \int_{0}^{t}\textrm{d}\tau_{1} \int_{0}^{t}\textrm{d}\tau_{2} \, Q_{c}(t-\tau_{1}) Q_{s}(t-\tau_{2}) N_{\alpha_{j}}(\tau_{1},\tau_{2}) \, ,
\end{align}
where $j=x,y$ and to compute $C_{j}(t)$, we have used the symmetry
$N_{\alpha_{j}}(\tau_{1},\tau_{2}) =
N_{\alpha_{j}}(\tau_{2},\tau_{1})$. Since we are interested in the
nonequilibrium steady state (NESS), we evaluate the quantities in
Eqs.~(\ref{011a})-(\ref{011c}) in the limit $t \rightarrow
\infty$. Details of the derivation can be found in
\ref{Appendix_Auxilliary}. Eventually, we find
\begin{align}
\label{29112021_0944a}
A_{j} 
& = \frac{\Gamma(1+\alpha_{j})}{4} \biggl[ \frac{2-u}{(1-u)^{\alpha_{j}}} + \frac{2+u}{(1+u)^{\alpha_{j}}} \biggr] \\
\label{29112021_0944b}
B_{j}
 & = \frac{\Gamma(1+\alpha_{j})}{4} \biggl[ \frac{u}{(1-u)^{\alpha_{j}}} - \frac{u}{(1+u)^{\alpha_{j}}} \biggr]  \\
\label{29112021_0944c}
C_{j} 
 & = \frac{\Gamma(1+\alpha_{j})}{4} \biggl[ \frac{1}{(1-u)^{\alpha_{j}}} - \frac{1}{(1+u)^{\alpha_{j}}} \biggr] \, .
\end{align}
where $A_j,B_j,C_j$ denote the limiting values of $A_j(t),B_j(t),C_j(t)$
in the limit $t\to \infty$.

One can notice that the asymptotic values of $A_j$, $B_j$ and $C_j$
are finite (exist) only for $|u| < 1$. This is the only case of
interest to us. When $|u| \ge 1$, the physics is very different since
the particle is not trapped, a fact which is reflected in the diverging
mean-squared displacements along both directions as $t \to \infty$.

\subsection{Probability density function and characteristic function}
\label{Position_proba}
Since the noise is Gaussian and the equations of motion are linear,
the probability density function $P(x, y; t)$ is fully determined by
the second moments. This is most concisely expressed using the
characteristic function
\begin{align}
\label{006} \nonumber
\Phi(t, \lambda_{x},\lambda_{y}; T_{x},T_{y}; \alpha_{x}, \alpha_{y}) & = \langle \textrm{e}^{ \im \lambda_{x} x(t) + \im \lambda_{y} y(t) } \rangle \\
& = \exp\Bigl[ -\frac{1}{2} \left(\langle x^{2}(t) \rangle\lambda_{x}^{2} + \langle y^{2}(t) \rangle \lambda_{y}^{2} + 2 \langle x(t)y(t) \rangle \lambda_{x} \lambda_{y} \right) \Bigr] \, ,
\end{align}
where $\langle x^{2}(t) \rangle$, $\langle y^{2}(t) \rangle$, and
$\langle x(t) y(t)\rangle$ are given by
Eqs.~(\ref{016a})-(\ref{016c}). The calculation of the characteristic
function Eq.~(\ref{006}) is performed in \ref{Appendix_Characteristic}.

The probability density function is then obtained as the inverse
Fourier transform of Eq.~(\ref{006}), which gives
\begin{equation}
	\label{017}
	P(x,y;t) = \frac{1}{2\pi\sqrt{ \langle x^{2}(t) \rangle \langle y^{2}(t) \rangle - \langle x(t)y(t) \rangle^{2}} } \exp\biggl[ - \frac{  \langle y^{2}(t) \rangle x^{2} +  \langle x^{2}(t) \rangle y^{2} - 2 \langle x(t)y(t) \rangle  x y }{ 2 ( \langle x^{2}(t) \rangle  \langle y^{2}(t) \rangle  - \langle x(t)y(t) \rangle^{2} ) } \biggr] \, .
\end{equation}
The normalization
$\int_{\mathbb{R}^{2}} \textrm{d}x \textrm{d}y \, P(x,y;t) = 1$ is
ensured since the condition
$\langle x^{2}(t) \rangle \langle y^{2}(t) \rangle > \langle x(t)y(t)
\rangle^{2}$ is satisfied as long as the moments remain finite.

Integrating Eq.~(\ref{017}) over one variable, the marginal
distributions $P(x;t)$ and $P(y;t)$ take the expected form
\begin{equation}
  P(x;t)  = \frac{1}{\sqrt{2\pi  \langle x^{2}(t) \rangle}} \textrm{e}^{-\frac{x^{2}}{2 \langle x^{2}(t) \rangle}};\quad
  P(y;t ) = \frac{1}{\sqrt{2\pi  \langle y^{2}(t) \rangle}} \textrm{e}^{-\frac{y^{2}}{2 \langle y^{2}(t) \rangle}}\,.
\end{equation}

\subsection{Quantitative analysis of the effect of the FGN}
\label{Position_quantitative}
We discuss here the effect of the FGN on the position moments in the
special cases of Model 1 and 2.

\begin{figure}[htbp]
\centering
\includegraphics[width=\textwidth]{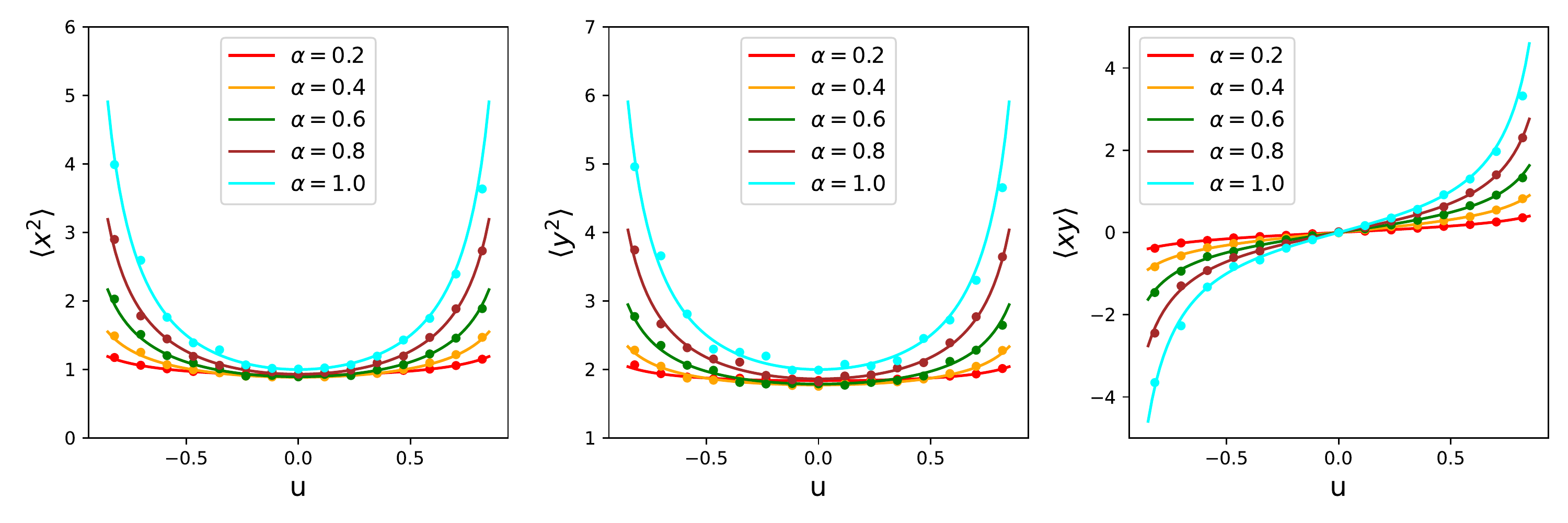}\\
\includegraphics[width=\textwidth]{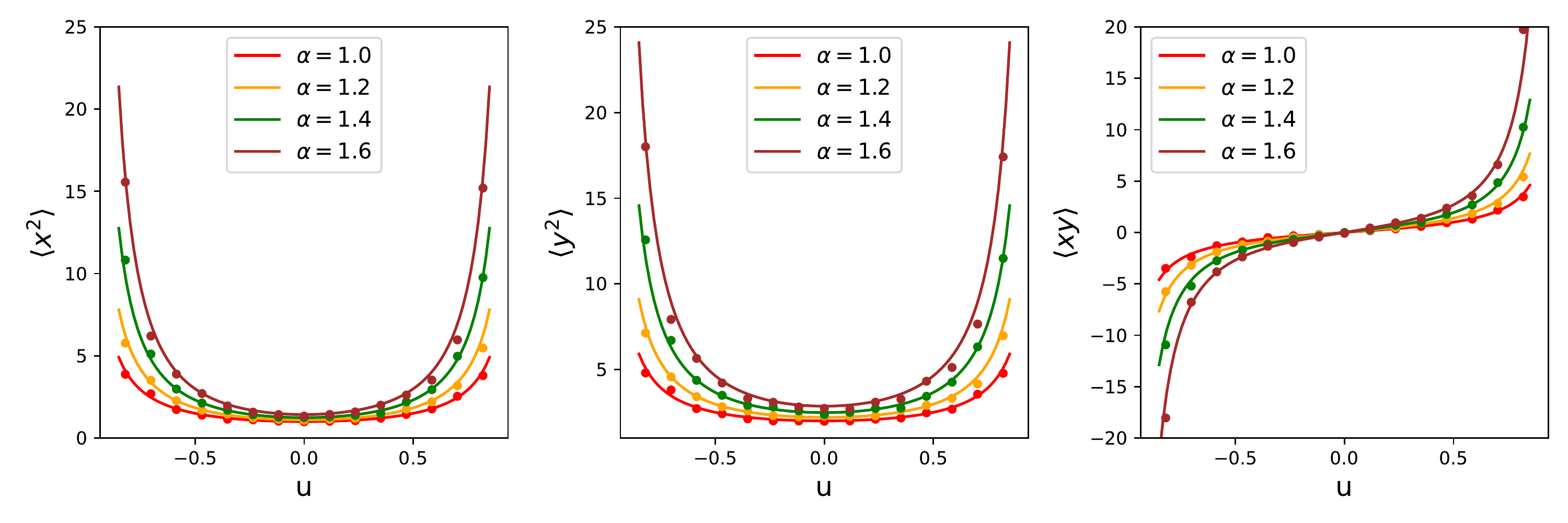}
\caption{Model $1$: $\langle x^2\rangle$, $\langle y^2\rangle$ and
  $\langle xy\rangle$ versus $u$ for varying $\alpha$. $T_x=1$ and
  $T_{y}=2$. We separated the sub-diffusive case (top) and the
  super-diffusive case (bottom) for readability. Full curves correspond
  to the exact results and dots to the simulation data.}
\label{fig:varmode1}
\end{figure}

In Model 1, which has $\alpha_x=\alpha_y$ and $T_x\neq T_y$, the second
moments Eq.~(\ref{016a})-(\ref{016c}) in the steady state simplify to
\begin{align}
  \langle x^2\rangle &=\frac{\Gamma(1+\alpha)}{4}\left[ 2\left(\frac{1}{(1-u)^\alpha}+\frac{1}{(1+u)^\alpha}\right) T_x +u \left(\frac{1}{(1-u)^\alpha}-\frac{1}{(1+u)^\alpha}\right)(T_x-T_y)\right] \nonumber\\
  \langle x y\rangle &= \frac{\Gamma(1+\alpha)}{4}\left(\frac{1}{(1-u)^\alpha}-\frac{1}{(1+u)^\alpha}\right)(T_x-T_y)
\label{momentsModel1}
\end{align}
and $\langle y^2\rangle$ is obtained by interchanging $T_x$ and $T_y$
in the expression for $\langle x^2\rangle$. Putting $\alpha=1$, one
recovers the results of
Ref\cite{Dotsenko2013,Mancois2018}. Fig.~\ref{fig:varmode1} shows the
prediction of Eq.~(\ref{momentsModel1}) for varying
$\alpha$. Comparing with numerical simulations confirms that the
predictions are exact. We observe that, varying $\alpha$, the shape of
the curves is not affected. However, quantitatively, $\alpha$ has a
strong effect. In particular, the dependence on $u$ becomes much
steeper when $\alpha$ increases, with the amplitude of all quantities
increasing by more than an order of magnitude for all $|u|>0.5$ when
increasing $\alpha$ from $0.2$ to $1.6$.

When $u=0$, we find, as expected, that the two components decouple,
{\it i.e.} $\langle x y\rangle =0$. The variance becomes
$\langle x^2\rangle=\Gamma(1+\alpha)T_x$ so that we can interpret
$\Gamma(1+\alpha)T_x\equiv T_{\rm eff,x}$ as an effective temperature
and the marginal distribution $P(x)\propto e^{-x^2/(2 T_{\rm eff,x})}$
as an effective Boltzmann distribution although the system remains out
of equilibrium as discussed in Sec.~\ref{Model}.



\begin{figure}[htbp]
  \centering
  \includegraphics[width=\textwidth]{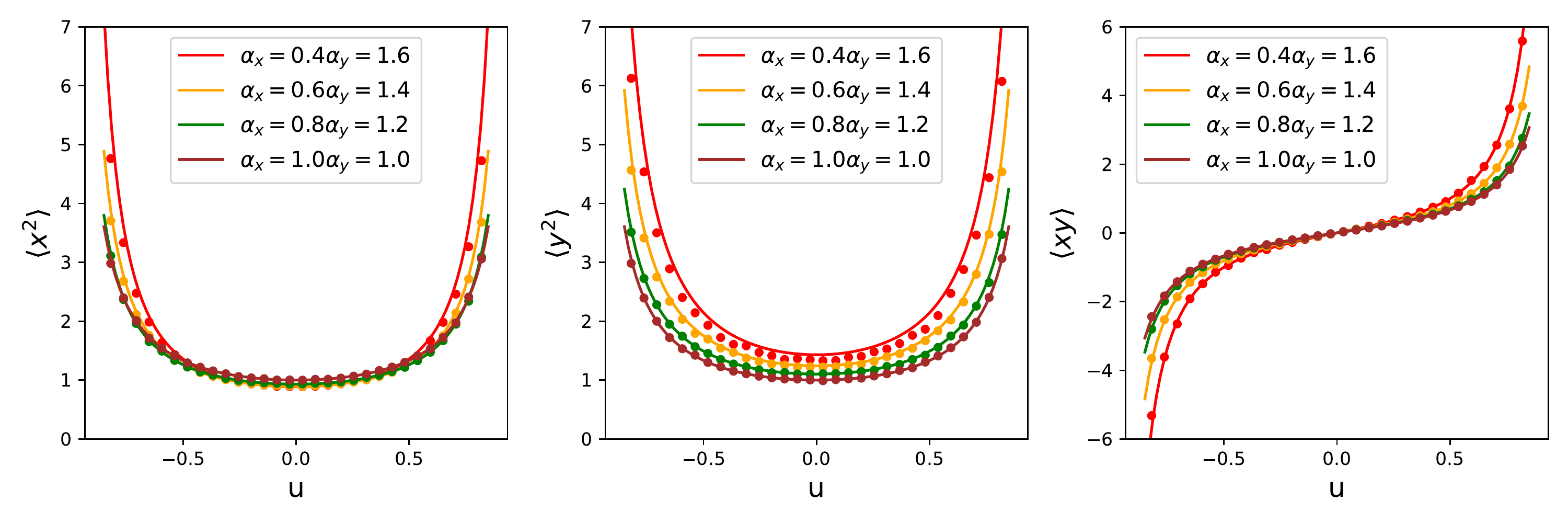}
  \caption{Model $2$: $\langle x^2\rangle$, $\langle y^2\rangle$ and
    $\langle xy\rangle$ versus $u$ with $T_x=T_y=1$ and varying
    difference in the exponent $\alpha_x=2-\alpha_y$. Full curves
    correspond to the exact results and dots to the simulation data.}
  \label{fig:varmode2}
\end{figure}

For Model 2, which has $T_x=T_y$ and $\alpha_x\neq \alpha_y$, Figure~\ref{fig:varmode2} shows the moments $\langle x^2\rangle$,
$\langle y^2\rangle$ and $\langle xy\rangle$ as functions of the
anisotropy parameter $u$ when increasing the difference in exponent
while keeping $\alpha_x=2-\alpha_y$. We see that increasing the
difference in exponent from $0$ to $1.2$, the amplitude of all
quantities increases, more significantly as $|u|$ gets larger,
although the effect remains moderate.

\section{Currents and rotation}
\label{Rotation}
A NESS is characterized by the existence of a probability current. We
first compute this steady-state current in Sec.~\ref{SScurrent}. We
then use it to characterize the rotational motion of our gyrator
through the angular velocity in Sec.~\ref{angularV} and the curl of
the current in Sec.~\ref{curlC}.

\subsection{Steady-state current}
\label{SScurrent}
Computing the current for our FBG involves subtleties due to the fact
that for the noises with long-ranged temporal correlations, we cannot
write a closed-form differential equation for the probability density
function which would generalize the Fokker-Planck equation that can be
written when $\alpha=1$ and give us an expression of the current.  We
circumvent here this fundamental difficulty by starting directly from
the definition of the current
$\textbf{J}(x,y,t) = (J_{x}(x,y,t),J_{y}(x,y,t))$ as an
ensemble-average over noise realizations \footnote{See e.g.,
  \cite{Dotsenko2019,Bae2021} for a definition involving also
  time-averages.}
\begin{eqnarray}
	\label{021}
	J_{x}(x, y; t) & = & \langle \dot{x}(t) \delta(x-x(t)) \delta(y-y(t)) \rangle \\
	\label{022}
	J_{y}(x, y; t) & = & \langle \dot{y}(t) \delta(x-x(t)) \delta(y-y(t)) \rangle \, .
\end{eqnarray}
It is easy to check that this is a good definition of the current, in
the sense that it satisfies the continuity equation
$\partial_t P(x,y,t)+\nabla\cdot \textbf{J}(x,y,t)=0$.

Using the integral representation of the Dirac delta function, $J_{x}$
can be expressed as
\begin{equation}
	\label{030}
	J_{x}(x, y; t) = \int_{\mathbb{R}^{2}}\frac{\textrm{d}\lambda_{x}\textrm{d}\lambda_{y}}{(2\pi)^{2}} \textrm{e}^{ - \im x \lambda_{x} - \im y \lambda_{y} } \langle \dot{x}(t) \textrm{e}^{ \im \lambda_{x} x(t) + \im \lambda_{y} y(t) } \rangle \, .
\end{equation}
which becomes, using the equation of motion Eq.~(\ref{001a}),
\begin{equation}
	\label{031}
	J_{x}(x, y; t) = \int_{\mathbb{R}^{2}}\frac{\textrm{d}\lambda_{x}\textrm{d}\lambda_{y}}{(2\pi)^{2}} \textrm{e}^{ - \im x \lambda_{x} - \im y \lambda_{y} } \langle \bigl[ - x(t) + u y(t) + \xi_{x}(t) \Bigr] \textrm{e}^{ \im \lambda_{x} x(t) + \im \lambda_{y} y(t) } \rangle \, .
\end{equation}
The average containing the first two terms on the r.h.s. can be
computed by using the characteristic function
\begin{equation}
	\label{032}
	\langle x(t) \textrm{e}^{ \im \lambda_{x}x(t) + \im \lambda_{y}y(t) } \rangle   =  - \im \partial_{\lambda_{x}} \Phi(t, \lambda_{x},\lambda_{y}) \, ;\quad
	\langle y(t) \textrm{e}^{ \im \lambda_{x}x(t) + \im \lambda_{y}y(t) } \rangle   =  - \im \partial_{\lambda_{y}} \Phi(t, \lambda_{x},\lambda_{y}) \, .
\end{equation}
with $\Phi(t,\lambda_{x},\lambda_{y})$ (we have omitted the dependence
on the other parameters for brevity) given in \ref{appendix_phi}. The
last term requires different characteristic functions
\begin{equation}
  \label{033}
	\Xi_{x}  = \langle \xi_{x}(t) \textrm{e}^{ \im \lambda_{x}x(t) + \im \lambda_{y}y(t) } \rangle; \qquad 
	\Xi_{y}  = \langle \xi_{y}(t) \textrm{e}^{ \im \lambda_{x}x(t) + \im \lambda_{y}y(t) } \rangle \, ,
\end{equation}
which are given in \ref{appendix_xi}.

Putting everything together (see details in \ref{appendix_Current}) we
arrive at the following expression for the current in steady state
\begin{align}
  \label{eq:currents}
	J_{x}(x, y)  &= - P(x,y) \partial_{x}U(x,y) - \Gamma(1+\alpha_x) T_{x} \partial_{x}P(x,y)\\
	J_{y}(x, y)  &= - P(x,y) \partial_{y}U(x,y) - \Gamma(1+\alpha_y) T_{y} \partial_{y}P(x,y) \, .
\end{align}
We recognize the effective temperatures
$T_{\rm eff,x/y}=\Gamma(1+\alpha_{x/y})T_{x/y}$ introduced in
Sec.~\ref{Position_quantitative} which control the probability
distribution at $u=0$ when the two components are decoupled. In
addition, we see from Eq.~(\ref{eq:currents}) that for any $u$ the
relation between current and probability is the same as the one that
would be obtained for a standard Brownian gyrator driven by Gaussian
white noises at the effective temperatures $T_{\rm eff,x/y}$.

\begin{figure}[htbp]
	\centering
	\includegraphics[scale=0.6]{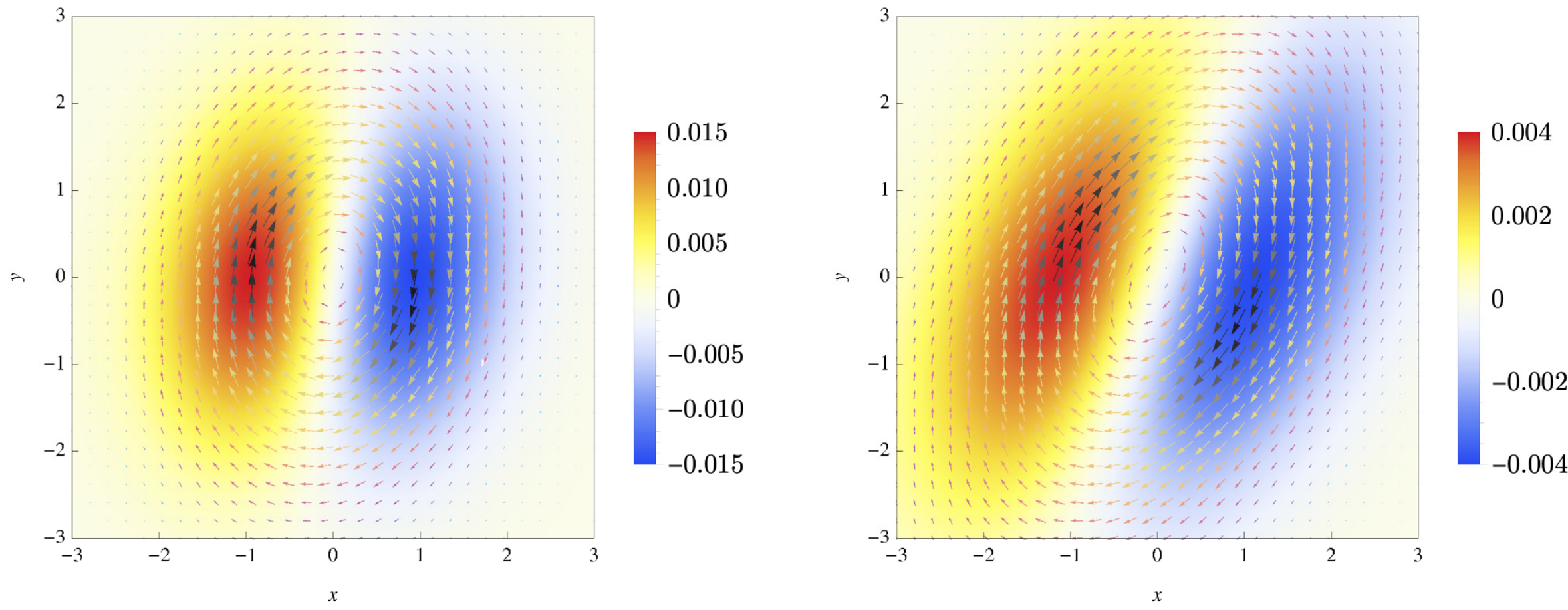}
	\caption{Model 1: Current field with arrows indicating the
          direction and the color map indicating the amplitude. Left
          panel: $\alpha=0.5$. Right panel: $\alpha=1.5$. In both
          panels $T_{x}=1$, $T_{y}=2$, $u=0.3$.}
	\label{fig_m2_1ter}
\end{figure}

The current field is shown in Fig.~\ref{fig_m2_1ter} for the special
case of Model 1 with $T_{x}=1$, $T_{y}=2$ and $u=0.3$ and two values
of $\alpha$. Interestingly, one observes that when $\alpha$ becomes
larger, the symmetry axis of the current rotates.

\subsection{Angular velocity}
  \label{angularV}
One can use the expression Eq.~(\ref{eq:currents}) for the current to
compute the steady-state angular velocity
\begin{equation}
  \label{eq:omega-J-def}
 \langle {\bm \omega} \rangle =\int d^2{\bf r}\,\,\frac{1}{r^2}{\bf r}\land {\bf J}
\end{equation}
Since the current vector belongs to the $xy$ plane, we write
$\langle {\bm \omega} \rangle=\langle \omega \rangle {\bf \hat{z}}$
and, after straightforward algebra, we obtain the explicit, although
lengthy, expression:
\begin{align}\label{eq:omega}
\langle \omega \rangle &=\Bigg(T_x\Gamma(1+\alpha_x)F(u,\alpha_x)-T_y\Gamma(1+\alpha_y)F(u,\alpha_y)\Bigg)\nonumber\\	
&\quad \Bigg{/}\Bigg(H(\alpha_x,u) (T_x\Gamma(1+\alpha_x))^2 +H(\alpha_y,u) (T_y\Gamma(1+\alpha_y))^2\nonumber\\
&\qquad +2M(\alpha_x,\alpha_y,u)T_xT_y\Gamma(1+\alpha_x)\Gamma(1+\alpha_y)\Bigg)^{1/2}	\,,
\end{align}
where 
\begin{align}
F(u,\alpha)&=(1+u)^{2-\alpha} -(1-u)^{2-\alpha}\\
  H(u,\alpha)&=-(1+u)^{2-2\alpha} -(1-u)^{2-2\alpha}+2(1+u^2)(1-u^2)^{-\alpha} \\
  M(\alpha_x,\alpha_y,u)&=(1+u)^{2-\alpha_x-\alpha_y} +(1-u)^{2-\alpha-\alpha_y}\nonumber\\
	&+(3-u^2)\left((1+u)^{-\alpha_x}(1-u)^{-\alpha_y}+(1-u)^{-\alpha_x}(1+u)^{-\alpha_y} \right)
\end{align}
As for the standard BG, one needs a double symmetry breaking
(geometric and in the noise) to observe a rotation. Indeed, one sees
from Eq.~(\ref{eq:omega}) that the angular velocity vanishes if either
$u=0$ or if $\alpha_x=\alpha_y$ and $T_x=T_y$.

On the contrary, as soon as both symmetry breakings are present, one
observes a rotation in general. This is most clearly seen by
performing a series expansion $\langle \omega \rangle$ in powers of
$u$ to simplify Eq.~(\ref{eq:omega}) to
\begin{equation}\label{eq:omega2}
	\langle \omega \rangle =\frac{T_x\Gamma(1+\alpha_x)(2-\alpha_x)-T_y\Gamma(1+\alpha_y)(2-\alpha_y)}{2\sqrt{T_xT_y\Gamma(1+\alpha_x)\Gamma(1+\alpha_y)}}u+O(u^3)
\end{equation}
At this order, the second symmetry breaking occurs when  
 \begin{equation}\label{eq:symbreak}
 T_y\Gamma(1+\alpha_y)(2-\alpha_y)\neq T_x\Gamma(1+\alpha_x)(2-\alpha_x).
\end{equation}
This result encompasses the cases of both Models 1 and 2, showing that
rotation is generated by having either different temperatures or
different exponents. Conversely, rotation can be diminished, even with
different noises, by choosing the parameters so that
Eq.~(\ref{eq:symbreak}) becomes an equality.

\begin{figure}[htbp]
	\centering
	\includegraphics[width=0.48\textwidth]{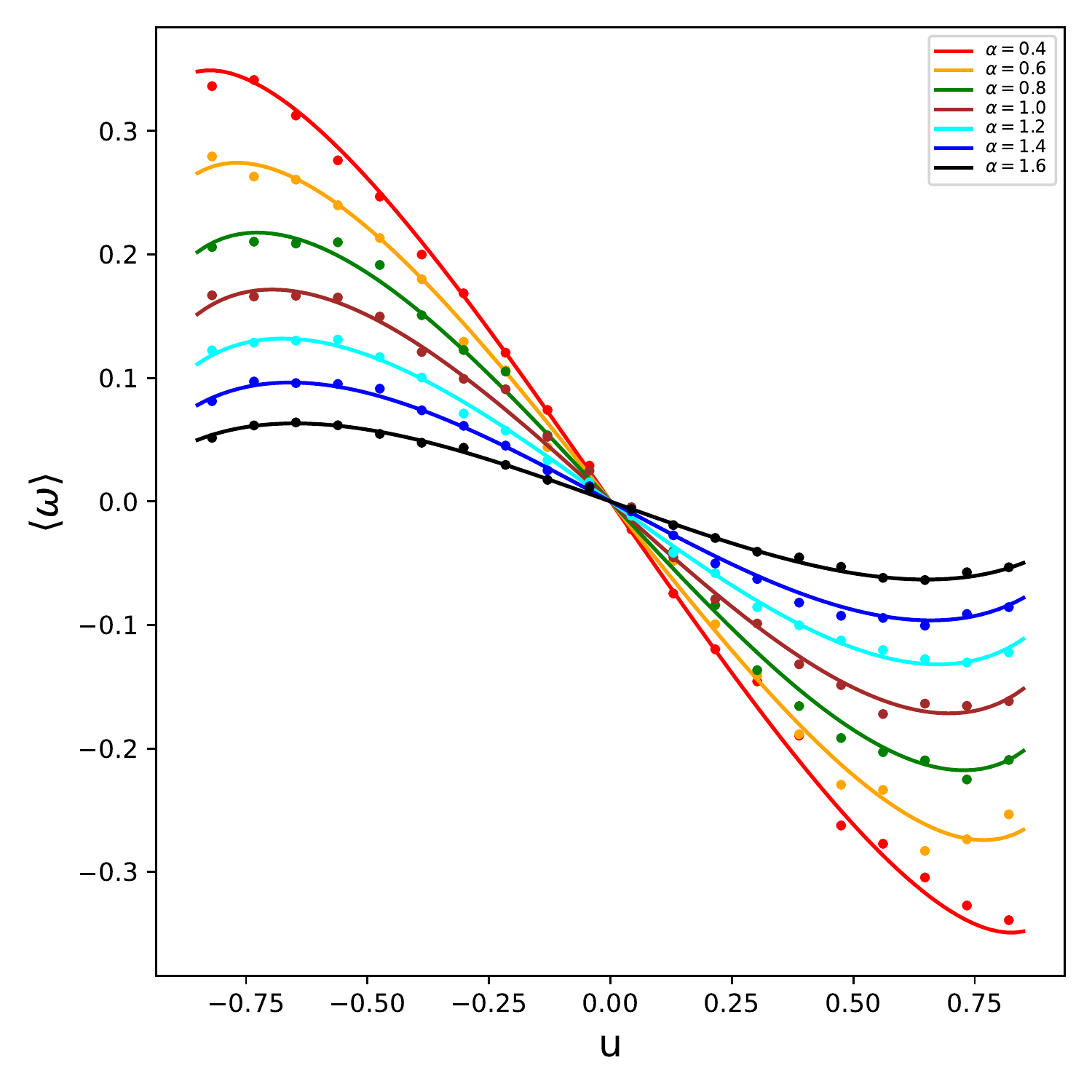}
	\includegraphics[width=0.48\textwidth]{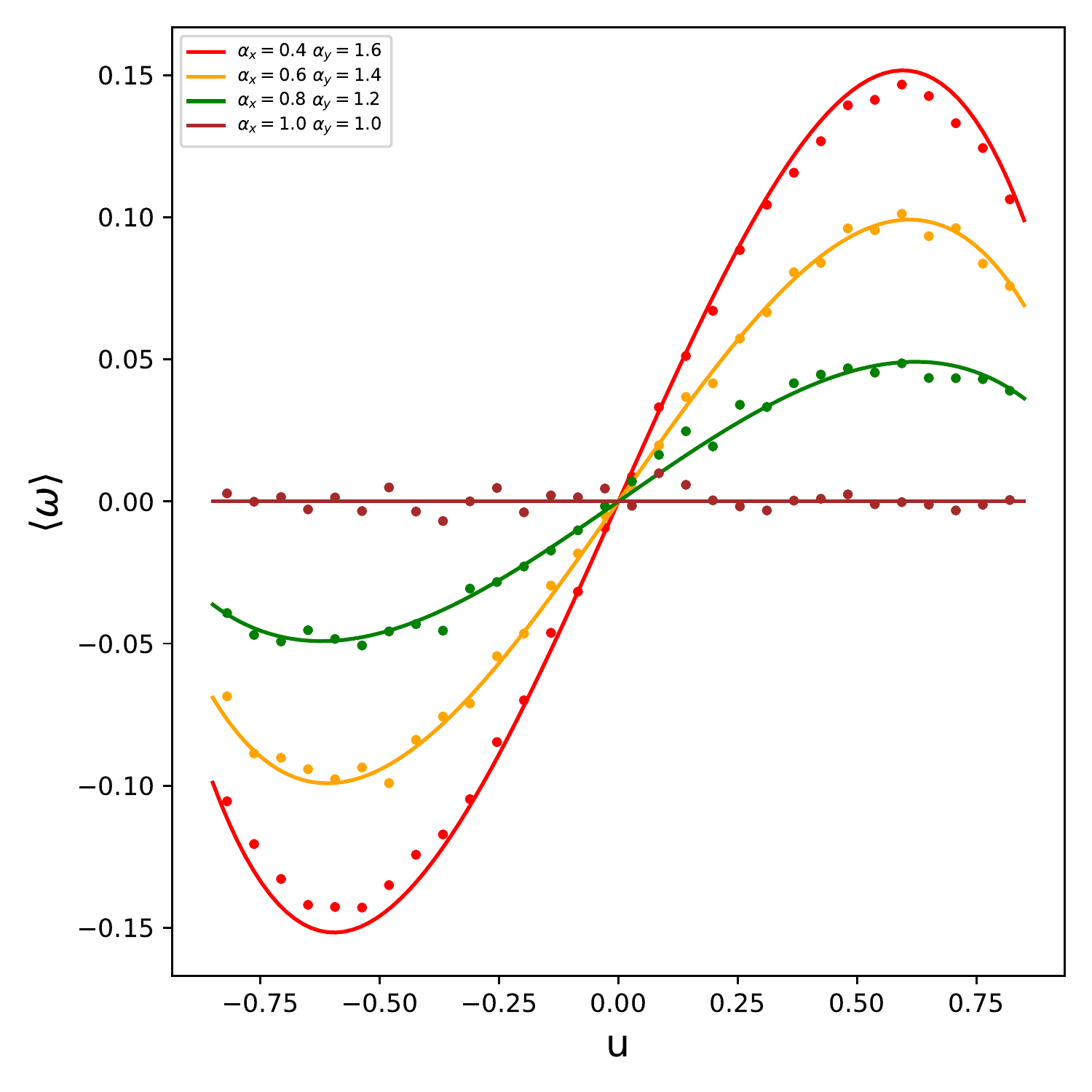}
	\caption{Mean angular velocity as function of the anisotropy
          parameter $u$. Left: Model 1 with $T_x=1$ and
          $T_{y}=2$. Right: Model 2 with $T_x=T_y=1$.}
	\label{fig:angular1}
\end{figure}

Let us now discuss the predictions of Eq.~(\ref{eq:omega}) and compare
with numerical measurements. In practice, we do not use
Eq.~(\ref{eq:omega-J-def}) to compute the angular velocity in
simulations but rather compute the time derivative of the polar
angle. The mean of $\omega$ is displayed in Fig.~\ref{fig:angular1}
for Model 1 and Model 2. For Model 1, we observe that
$\langle \omega\rangle$ becomes larger as the asymmetry $|u|$
increases and as $\alpha$ decreases. For Model 2,
$\langle \omega\rangle$ shows two extrema with $|u|$ and increases in
amplitude when the difference between the exponents increases.

\subsection{Curl of the current}
\label{curlC}

The rotational motion of the particle in the NESS can also be
characterized by the curl of the current ${\bf \nabla} \land {\bf J}$ which
can be computed from Eq.~(\ref{eq:currents}). In particular, let us
give here its expression at the origin in the case of Model 1
\begin{equation}
	\label{curl0}
{\bf \nabla}  \land {\bf J}(0,0) = \frac{T_{x}^{2}-T_{y}^{2}}{4\pi(T_{x}T_{y})^{3/2}} \frac{(1-u^{2})^{\alpha/2}}{\Gamma(1+\alpha)} \frac{G_{+}(2u,u,\alpha)}{[1+G_{-}(1+u^{2},u,\alpha)\Delta^{2}]^{3/2}}
\end{equation}
where
\begin{equation}
G_{\pm}(x,u,\alpha) = \frac{x}{2} + \frac{1-u^{2}}{4} \biggl[ - \left( \frac{1+u}{1-u} \right)^{\alpha-1} \pm \left( \frac{1-u}{1+u} \right)^{\alpha-1} \biggr]; \quad \Delta = \frac{T_{y}-T_{x}}{2\sqrt{T_{x}T_{y}}} \, .
\end{equation}
Again one can see the double symmetry breaking that is necessary to
obtain a rotational motion since ${\bf \nabla} \land {\bf J}(0,0)$ in
Eq.~(\ref{curl0}) vanishes when either $T_x=T_y$ or
$u=0$. 
The known result for the Brownian gyrator~\cite{Dotsenko2013},
\begin{equation}
{\bf \nabla}  \land {\bf J}_{\rm BG}(0,0) = \frac{T_{x}^{2}-T_{y}^{2}}{4\pi(T_{x}T_{y})^{3/2}} \frac{u\sqrt{1-u^{2}}}{(1+u^{2}\Delta^{2})^{3/2}} \, ,
\end{equation}
is easily retrieved by observing that $G_{+}(2u,u,1)=u$ and
$G_{-}(1+u^{2},u,1)=u^{2}$.

\begin{figure}[htbp]
	\centering
	\includegraphics[width=0.48\textwidth]{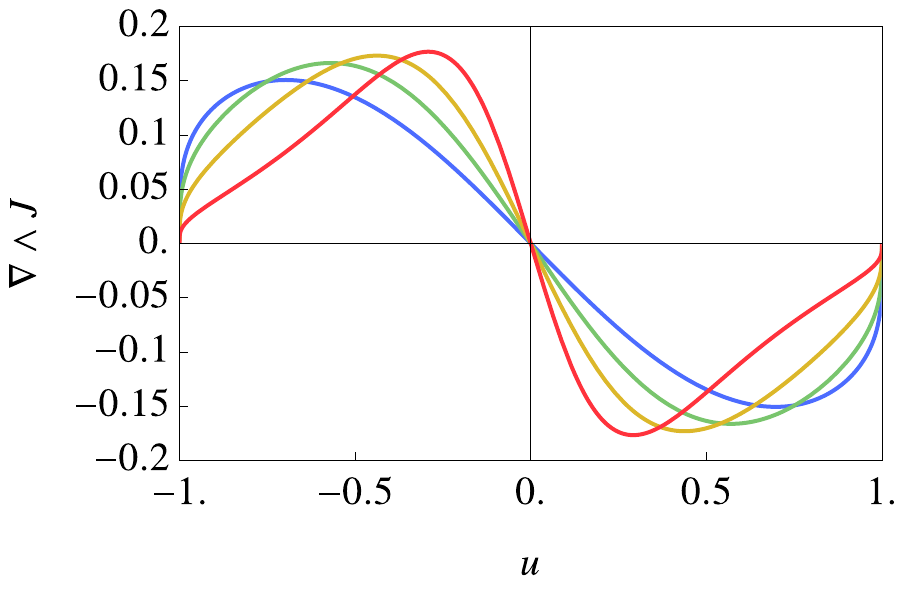}
	\includegraphics[width=0.48\textwidth]{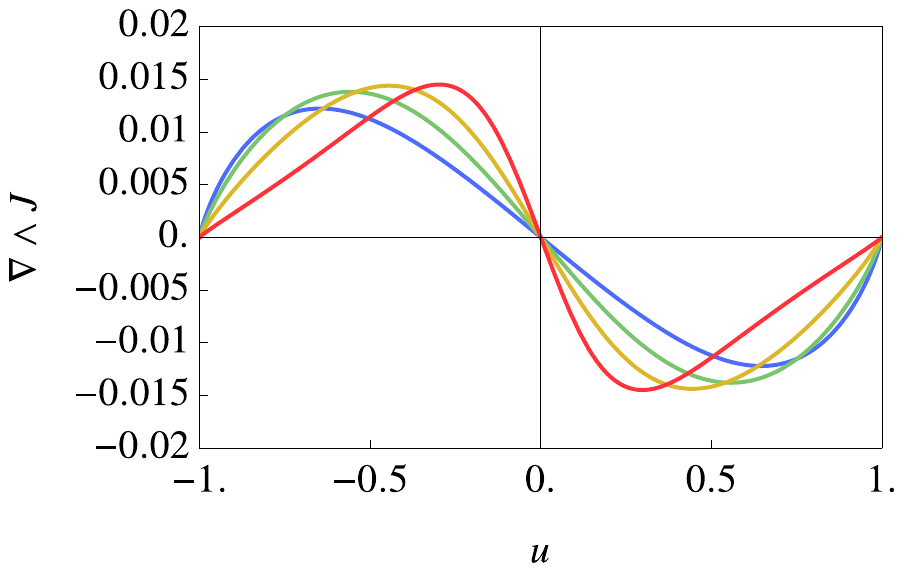}
	\caption{Model 1: ${\bf \nabla} \land {\bf J}(0,0)$ as a
          function of $u$ for $T_{y}$ taking values $5$ (blue), $10$
          (green), $20$ (dark yellow), and $50$ (red). Left panel:
          $\alpha=0.25$. Right panel: $\alpha=1.75$. $T_x=1$.}
	\label{fig_m2_1}
\end{figure}

\begin{figure}[htbp]
\centering
\includegraphics[width=\textwidth]{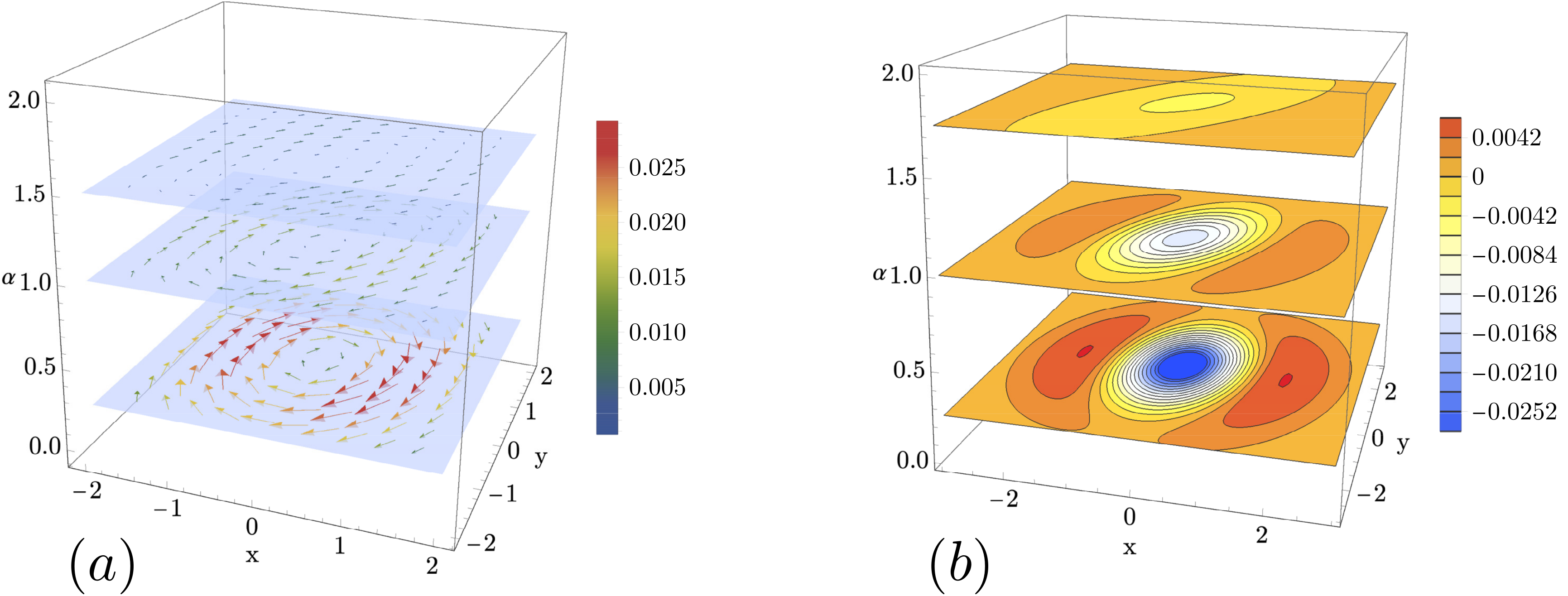}
\caption{Model 1: Panel (a) the current field $\mathbb{J}(x,y)$ in the
  steady state. The vertical slices are taken at $\alpha=0.25$,
  $\alpha=1$ and $\alpha=1.75$. Panel (b): the $z$-component of
  $\nabla\wedge\textbf{J}(x,y)$ at the same values of $\alpha$. For
  both panels: $T_{x}=1$, $T_{y}=2$, $u=0.5$}
\label{fig_m2_1bis}
\end{figure}

Fig.~\ref{fig_m2_1} shows the curl at the origin for various
temperatures and two exponents. One observes that the extrema depend
little on the temperature but strongly on $\alpha$, increasing by an
order of magnitude when going from $\alpha=1.75$ to $\alpha=0.25$. The
latter trend is also observed on the current field and its curl at
arbitrary positions, displayed in Fig.~\ref{fig_m2_1bis}.

\begin{figure}[htbp]
	\centering
	\includegraphics[scale=0.8]{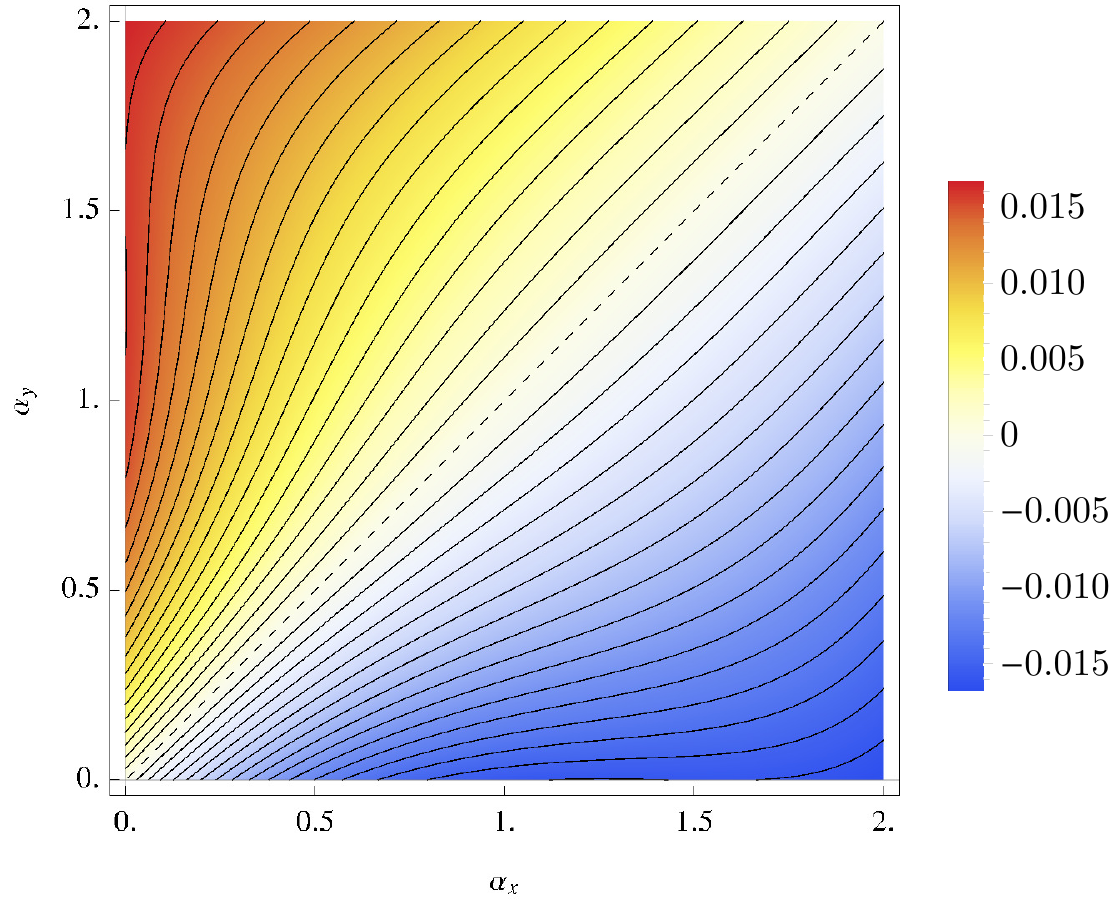}
	\caption{Model 2: Density plot of ${\bf \nabla} \land {\bf
            J}(0,0)$ as a function of $\alpha_{x}$, $\alpha_{y}$ with
          $T=1$ and $u=0.1$. The curl vanishes for $\alpha_x=\alpha_y$ (dashed
          line).}
	\label{fig_m2_2}
\end{figure}

In the case of Model 2, unsurprisingly, the curl at the origin
increases when the difference between the exponents increase, as
illustrated in Fig.~\ref{fig_m2_2}.

\section{Conclusion}
\label{conc}

We have considered the two-dimensional motion of a particle in a
harmonic potential. The particle undergoes Brownian motion in the
overdamped limit under the action of different fractional Gaussian
noises along the two spatial directions. As such it is out of
equilibrium and display a non-vanishing current in steady state. Most
strikingly, the particle rotates around the origin. This model can be
seen as a minimal model of a nonequilibrium steady state driven by
long-range correlated noise that can be tuned to be sub-diffusive or
super-diffusive. We have computed analytically the full position
probability distribution and the current in steady state, from which
we deduced the angular velocity and the curl of the current
characterizing the rotational motion of the particle. Numerical
simulations confirmed these derivations in two special cases where the
rotation is driven either by different amplitudes of the noise or by
different exponents. The effect of the fractional noise, compared to the
white noise case $\alpha=1$ is rather paradoxical: When decreasing
$\alpha$ the position distribution becomes more and more insensitive
to the asymmetry $u$ but at the same time the current and angular
velocity increase. For large values of $\alpha$, the particle
probability becomes much more spread and sensitive to the asymmetry
but rotates slower around the origin.

Of course, the properties of the nonequilibrium steady-state that we
have determined here depend on the details of the dynamics. For
example, we expect that using a generalized Langevin equation as in
Ref.~\cite{Nascimento2021} but including power-law correlations
instead of exponential ones will lead to a different steady state
compared to our result even though the free dynamics of each component
is exactly the same in both models (see {\it e.g.}
\cite{Jeon2012,Jeon2012a,Squarcini2022}). Long-ranged correlated
non-Gaussian noise could also yield different results. However, there
is hope that some universality is at play and that, qualitatively, our
conclusions capture the effect of long-ranged time correlations beyond
a  specific model. It would be interesting to question this
universality by studying other models in the future or even by
comparing to experimental measurements.

\section*{acknowledgments}
\label{conc}
A.S. acknowledges the warm hospitality of LPTMC, Sorbonne Universit\'e during his research stay in October- December 2021.

 

\appendix

\section{Auxiliary functions}
\label{Appendix_Auxilliary}
The calculation of $A_{j}(t)$, $B_{j}(t)$, and $C_{j}(t)$ involves integrals of the following form
\begin{equation}
	\label{app_001}
	g_{\alpha}(t) = \int_{0}^{t}\textrm{d}t_{1} \int_{0}^{t}\textrm{d}t_{2} \, \omega_{1}(t-t_{1}) \omega_{2}(t-t_{2}) N_{\alpha}(t_{1},t_{2}) \, ,
\end{equation}
the choice of the kernel $\omega_{j}$ depending on the specific function one is considering. We use the fact that
\begin{equation}
	\label{app_002}
	N_{\alpha}(t_{1},t_{2}) = \partial_{t_{1}}\partial_{t_{2}} \mathcal{G}_{\alpha}(t_{1},t_{2}) \, ,
\end{equation}
where
\begin{equation}
	\label{app_003}
	\mathcal{G}_{\alpha}(t_{1},t_{2}) =  \left( t_{1}^{\alpha} + t_{2}^{\alpha} - |t_{1}-t_{2}|^{\alpha} \right)
\end{equation}
is the autocorrelation function in free space. Integrating by parts twice we find an expression for $g_{\alpha}(t)$ 
\begin{align}
	\label{app_004} \nonumber
	g_{\alpha}(t) & = \omega_{1}(0) \omega_{2}(0) \mathcal{G}_{\alpha}(t,t) + \omega_{1}(0) \int_{0}^{t}\textrm{d}t_{2} \, \omega_{2}^{\prime}(t-t_{2}) \mathcal{G}_{\alpha}(t,t_{2}) + \omega_{2}(0) \int_{0}^{t}\textrm{d}t_{1} \, \omega_{1}^{\prime}(t-t_{1}) \mathcal{G}_{\alpha}(t_{1},t) \\
	& + \int_{0}^{t}\textrm{d}t_{1} \int_{0}^{t}\textrm{d}t_{2} \, \omega_{1}^{\prime}(t-t_{1}) \omega_{2}^{\prime}(t-t_{2}) \mathcal{G}_{\alpha}(t_{1},t_{2}) \, ,
\end{align}
where $\omega^{\prime}(u) \equiv \textrm{d}\omega(u)/\textrm{d} u$.

\subsection{The function $A_{j}(t)$}
For the function $A_{j}(t)$ the kernel is given by
$\omega_{1}(t)=\omega_{2}(t)=Q_{c}(t)$. The first term in the right
hand side of Eq.~(\ref{app_004}) yields the contribution
$2 t^{\alpha}$ because $Q_{c}(0)=1$. The terms involving a single
integration are identical and their contribution is denoted
$A_{j}^{(\rm I)}(t)$. Then, the contribution stemming from the term
involving a double integral is denoted $A_{j}^{(\rm II)}(t)$. Hence,
\begin{equation}
	\label{app_007}
	A_{j}(t) = 2 t^{\alpha} + A_{j}^{(\rm I)}(t) + A_{j}^{(\rm II)}(t) \, ,
\end{equation}
with
\begin{equation}
	\label{app_008}
	A_{j}^{(\rm I)}(t) = 2\int_{0}^{t}\textrm{d}t_{1} \, Q_{c}^{\prime}(t-t_{1}) \mathcal{G}_{\alpha}(t,t_{1}) \, .
\end{equation}
Recalling that $Q_{c}^{\prime}(t) = [ u \sinh(ut) -  \cosh(ut) ]e^{-t} $, the change of variables $t-t_{1}=z t$ gives
	\begin{equation}
	\label{app_009}
	A_{j}^{(\rm I)}(t) = 2  t^{1+\alpha} \int_{0}^{1}\textrm{d}z \, \Bigl[ u \sinh(utz) -  \cosh(utz) \Bigr] \exp(-tz) \bigl[ 1 + (1-z)^{\alpha} - z^{\alpha} \bigr] \, .
\end{equation}
It is convenient to express the integral in terms of the function
\begin{equation}
	\label{app_010}
	F(q,\alpha) = \int_{0}^{1}\textrm{d}z \, \exp(-qz) \bigl[ 1 + (1-z)^{\alpha} - z^{\alpha} \bigr] \, .
\end{equation}
Therefore
\begin{equation}
	\label{app_011}
	A_{j}^{(\rm I)}(t) = -  t^{1+\alpha} \biggl[ (1-u) F((1-u)t,\alpha) + (1+u) F((1+u)t,\alpha) \biggr] \, .
\end{equation}

Since we are interested in $\lim_{t \rightarrow \infty} A_{j}(t)$, it is sufficient to consider the infinite-$q$ limit of the function $F(q,\alpha)$ for positive $q$. In such a limit we can use the asymptotic expansion
\begin{equation}
	\label{app_012}
	F(q,\alpha) = \frac{2}{q} - \frac{\Gamma(1+\alpha)}{q^{1+\alpha}} - \frac{\alpha}{q^{2}} + O(q^{-3}) \, , \qquad q \rightarrow \infty \, .
\end{equation}
Taking $0<u<1$, it follows that
\begin{equation}
	\label{app_013}
	A_{j}^{(\rm I)}(t) = - 4   t^{\alpha} + \frac{2\alpha  }{1-u^{2}} t^{\alpha-1} +   \Gamma(1+\alpha) \biggl[ \frac{1}{(1-u)^{\alpha}} + \frac{1}{(1+u)^{\alpha}} \biggr] + O(t^{\alpha-2}) \, .
\end{equation}
Let us consider now the contribution from the double integral
\begin{equation}
	\label{app_014}
	A_{j}^{(\rm II)}(t) = \int_{0}^{t}\textrm{d}t_{1} \int_{0}^{t}\textrm{d}t_{2} \, Q_{c}^{\prime}(t-t_{1}) Q_{c}^{\prime}(t-t_{2}) \mathcal{G}_{\alpha}(t_{1},t_{2}) \, .
\end{equation}
We further split $A_{j}^{(\rm II)}(t)$ into a contribution due to the
terms $t_{1}^{\alpha}+t_{2}^{\alpha}$ in
$\mathcal{G}_{\alpha}(t_{1},t_{1})$ and the remainder due to
$|t_{1}-t_{2}|^{\alpha}$. This splitting is denoted as follows
\begin{equation}
	\label{app_015}
	A_{j}^{(\rm II)}(t) = A_{j}^{(\rm II,1)}(t) + A_{j}^{(\rm II,2)}(t) \, ,
\end{equation}
with
\begin{equation}
	\label{app_016}
	A_{j}^{(\rm II,1)}(t) =  \int_{0}^{t}\textrm{d}t_{1} \int_{0}^{t}\textrm{d}t_{2} \, Q_{c}^{\prime}(t-t_{1}) Q_{c}^{\prime}(t-t_{2}) \left( t_{1}^{\alpha} + t_{2}^{\alpha} \right) \, ,
\end{equation}
and
\begin{equation}
	\label{app_017}
	A_{j}^{(\rm II,2)}(t) = -  \int_{0}^{t}\textrm{d}t_{1} \int_{0}^{t}\textrm{d}t_{2} \, Q_{c}^{\prime}(t-t_{1}) Q_{c}^{\prime}(t-t_{2}) |t_{1}-t_{2}|^{\alpha} \, ,
\end{equation}
The integrals in Eq.~(\ref{app_016}) are factorized, therefore
\begin{align}
	\label{app_018}
	A_{j}^{(\rm II,1)}(t) & = 2  \int_{0}^{t}\textrm{d}t_{1} \, t_{1}^{\alpha} Q_{c}^{\prime}(t-t_{1}) \int_{0}^{t} \textrm{d}t_{2} \, Q_{c}^{\prime}(t-t_{2}) \, .
\end{align}
The integration yields
\begin{equation}
	\label{app_019}
	A_{j}^{(\rm II,1)}(t) =  \left( -1 + \textrm{e}^{- t} \cosh (ut) \right) \biggl[ \phi_{\alpha}( (1-u)t ) + \phi_{\alpha}( (1+u)t ) \biggr] t^{\alpha} \, ,
\end{equation}
where
\begin{equation}
	\label{app_020}
	\phi_{\alpha}(z) = \textrm{e}^{-\im \pi \alpha} z^{-\alpha} \textrm{e}^{-z} (\Gamma (\alpha +1)-\Gamma (\alpha +1,-z)) \, .
\end{equation}
When $z \rightarrow + \infty$ we have
\begin{equation}
	\label{app_021}
	\phi_{\alpha}(z) = - 1 + \frac{\alpha}{z} + O(z^{-2}) \, , \qquad z \rightarrow + \infty \, .
\end{equation}
Focusing on those terms which in the large-$t$ expansion are not exponentially suppressed, we find
\begin{equation}
	\label{app_022}
	A_{j}^{(\rm II,1)}(t) = 2 t^{\alpha} - \frac{2 \alpha  }{1-u^{2}} t^{\alpha-1} + O(t^{\alpha-2}) \, .
\end{equation}
We can now consider the integral Eq.~(\ref{app_017}), which can be written in the form
\begin{equation}
	\label{app_023}
	A_{j}^{(\rm II,2)}(t) = -  \int_{0}^{t}\textrm{d}t_{1} \, Q_{c}^{\prime}(t_{1}) \int_{0}^{t}\textrm{d}t_{2} \,  Q_{c}^{\prime}(t_{2}) |t_{1}-t_{2}|^{\alpha} \, ,
\end{equation}
which follows by applying the change of variables
$t_{1} \rightarrow t-t_{1}$ and $t_{2} \rightarrow t-t_{2}$ in
Eq.~(\ref{app_017}). Since the correlation function
$\mathcal{G}_{\alpha}(t_{1},t_{2})$ is invariant under the interchange
of $t_{1}$ and $t_{2}$, we have
\begin{equation}
	\label{app_024}
	A_{j}^{(\rm II,2)}(t) = - 2  \int_{0}^{t}\textrm{d}t_{1} \, Q_{c}^{\prime}(t_{1}) \int_{0}^{t_{1}}\textrm{d}t_{2} \,  Q_{c}^{\prime}(t_{2}) (t_{1}-t_{2})^{\alpha} \, .
\end{equation}
Again, since we are interested in the large-$t$ behavior, we can
simply perform a Laplace transform with respect to $t$ and apply the
final value theorem. The Laplace transform is written as follows
\begin{equation}
	\label{app_025}
	\mathcal{L} \left( A_{j}^{(\rm II,2)}(t) ; t, s \right)= - \frac{2 }{s} \int_{0}^{\infty}\textrm{d}t_{1} \, \textrm{e}^{-st_{1}}Q_{c}^{\prime}(t_{1}) \int_{0}^{t_{1}}\textrm{d}t_{2} \,  Q_{c}^{\prime}(t_{2}) (t_{1}-t_{2})^{\alpha} \, ,
\end{equation}
where $s$ is the Laplace parameter. The integrations in Eq.~(\ref{app_025}) can be computed analytically and the corresponding result admits the following Taylor expansion for small $s$
\begin{equation}
	\label{app_026}
	\mathcal{L} \left( A_{j}^{(\rm II,2)}(t) ; t, s \right) = - \frac{2 }{s} \sum_{n=0}^{\infty} a_{n}(u,\alpha) s^{n} \, ,
\end{equation}
where $a_{n}(u,\alpha)$ are functions that can be determined
analytically. The final value theorem states that
\begin{equation}
	\label{app_027}
	\lim_{t \rightarrow \infty} A_{j}^{(\rm II,2)}(t) = \lim_{s \rightarrow 0} s\mathcal{L} \left( A_{j}^{(\rm II,2)}(t) ; t, s \right) \, ,
\end{equation}
and therefore
\begin{equation}
	\label{app_028}
	\lim_{t \rightarrow \infty} A_{j}^{(\rm II,2)}(t) = - 2  a_{0}(u,\alpha)\,\quad \text{with}\quad  a_{0}(u,\alpha) = \frac{\Gamma(1+\alpha)}{8} \biggl[ \frac{2+u}{(1-u)^{\alpha}} + \frac{2-u}{(1+u)^{\alpha}} \biggr] \,.
\end{equation}
Let us mention that it is also possible to invert the Laplace
transform in Eq.~(\ref{app_024}) and obtain the formal expansion for large
$t$, including also the subleading corrections for large
times. Collecting the results obtained so far, we find
\begin{equation}
	\label{app_030}
	\lim_{t \rightarrow \infty} A_{j}(t) = \frac{}{4} \Gamma(1+\alpha) \biggl[ \frac{2-u}{(1-u)^{\alpha}} + \frac{2+u}{(1+u)^{\alpha}} \biggr] \, .
\end{equation}
\subsection{The function $B_{j}(t)$}
The calculation of this function, which has kernels
$\omega_{1}(t)=\omega_{2}(t)=Q_{s}(t)$, turns out to be easier than
the calculation of $A_{j}(t)$ because $Q_{s}(t)$ vanishes at $t=0$. As
a result,
\begin{equation}
	\label{app_031}
	B_{j}(t) = \int_{0}^{t}\textrm{d}t_{1} \int_{0}^{t}\textrm{d}t_{2} \, Q_{s}^{\prime}(t-t_{1}) Q_{s}^{\prime}(t-t_{2}) \mathcal{G}_{\alpha}(t_{1},t_{2}) \, .
\end{equation}
We proceed by following the very same guidelines which we followed for
the calculation of $A_{j}(t)$. Using the same notation as before, we
split Eq.~(\ref{app_031}) by writing
\begin{equation}
	\label{app_032}
	B_{j}(t) = B_{j}^{(\rm II,1)}(t) + B_{j}^{(\rm II,2)}(t) \, ,
\end{equation}
where
\begin{align}
	\label{app_033}
	B_{j}^{(\rm II,1)}(t) & =  \int_{0}^{t}\textrm{d}t_{1} \int_{0}^{t}\textrm{d}t_{2} \, Q_{s}^{\prime}(t-t_{1}) Q_{s}^{\prime}(t-t_{2}) \left( t_{1}^{\alpha} + t_{2}^{\alpha} \right) \, , \\
	B_{j}^{(\rm II,2)}(t) & = -  \int_{0}^{t}\textrm{d}t_{1} \int_{0}^{t}\textrm{d}t_{2} \, Q_{s}^{\prime}(t-t_{1}) Q_{s}^{\prime}(t-t_{2}) |t_{1}-t_{2}|^{\alpha} \, .
\end{align}
The first integral gives
\begin{align}
	\label{app_034} \nonumber
	B_{j}^{(\rm II,1)}(t) & = 2  \int_{0}^{t}\textrm{d}t_{1} \, t_{1}^{\alpha} Q_{s}^{\prime}(t-t_{1}) \int_{0}^{t}\textrm{d}t_{2} \,  Q_{s}^{\prime}(t-t_{2}) \\ \nonumber
	& =   \sinh (ut) \textrm{e}^{-2t} \biggl[ (1-u)^{-\alpha } \textrm{e}^{-i \pi  \alpha +t u} (\Gamma (1+\alpha)-\Gamma (1+\alpha,-t (1-u))) \\ \nonumber
	& -  (1+u)^{-\alpha } \textrm{e}^{-i \pi  \alpha -t u} (\Gamma (1+\alpha)-\Gamma (1+\alpha ,-t (1+u))) \biggr] \\
	& =   \sinh (ut) \textrm{e}^{-t} t^{\alpha} \biggl[ \phi_{\alpha}((1-u)t) - \phi_{\alpha}((1+u)t) \biggr] \,.
\end{align}
When $t \rightarrow \infty$ the function $\phi_{\alpha}$ reduces to a constant while the overall $t$-dependence $\sim \exp(-(1-u)t)$ produces
\begin{equation}
	\label{app_035}
	\lim_{t \rightarrow \infty} B_{j}^{(\rm II,1)}(t) = 0 \, .
\end{equation}
The second integral in Eq.~(\ref{app_033}) is first written in the form
\begin{align}
	\label{app_036} \nonumber
	B_{j}^{(\rm II,2)}(t) & = -  \int_{0}^{t}\textrm{d}t_{1} \, Q_{s}^{\prime}(t-t_{1}) \int_{0}^{t}\textrm{d}t_{2} \, Q_{s}^{\prime}(t-t_{2}) |t_{1}-t_{2}|^{\alpha} \\ \nonumber
	& = -  \int_{0}^{t}\textrm{d}t_{1} \, Q_{s}^{\prime}(t_{1}) \int_{0}^{t}\textrm{d}t_{2} \, Q_{s}^{\prime}(t_{2}) |t_{1}-t_{2}|^{\alpha} \\
	& = - 2 \int_{0}^{t}\textrm{d}t_{1} \, Q_{s}^{\prime}(t_{1}) \int_{0}^{t_{1}}\textrm{d}t_{2} \, Q_{s}^{\prime}(t_{2}) (t_{1}-t_{2})^{\alpha} \, .
\end{align}
We apply the Laplace transform of both sides
\begin{equation}
	\mathcal{L}\left( B_{j}^{(\rm II,2)}(t) ; t, s \right) = - \frac{2 }{s} \int_{0}^{\infty}\textrm{d}t_{1} \, \textrm{e}^{-s t_{1}} Q_{s}^{\prime}(t-t_{1}) \int_{0}^{t_{1}}\textrm{d}t_{2} \, Q_{s}^{\prime}(t-t_{2}) |t_{1}-t_{2}|^{\alpha} \, ,
\end{equation}
and we get
\begin{equation}
	\mathcal{L}\left( B_{j}^{(\rm II,2)}(t) ; t, s \right) = - \frac{2 }{s} \sum_{n=0}^{\infty} b_{n}(u,\alpha) s^{n} \, ;\quad b_{0}(u,\alpha) = \frac{ u \Gamma (1+\alpha) }{ 8  } \bigl[ (1+u)^{-\alpha} - (1-u)^{-\alpha} \bigr]\, .
\end{equation}
The final value theorem then allows us to write
\begin{equation}
\lim_{t \rightarrow \infty} B_{j}(t)  = - 2 b_{0}(u,\alpha) =  \frac{ u \Gamma (1+\alpha) }{ 4  } \bigl[ (1-u)^{-\alpha} - (1+u)^{-\alpha} \bigr] \, .
\end{equation}

\subsection{The function $C_{j}(t)$}
Now using kernels $\omega_{1}(t)=Q_c(t)$ and $\omega_{2}(t)=Q_{s}(t)$,
we write as before
\begin{align}
	\label{} \nonumber
	C_{j}(t) & = C_{j}^{(\rm I)}(t) + C_{j}^{(\rm II)}(t) \, ,
\end{align}
where
\begin{align}
	\label{}
	C_{j}^{(\rm I)}(t) & = \int_{0}^{t}\textrm{d}t_{2} \, Q_{s}^{\prime}(t-t_{2}) \mathcal{G}_{\alpha}(t,t_{2}) \\
	C_{j}^{(\rm II)}(t) & = \int_{0}^{t}\textrm{d}t_{1} \int_{0}^{t}\textrm{d}t_{2} \, Q_{c}^{\prime}(t-t_{1}) Q_{s}^{\prime}(t-t_{2}) \mathcal{G}_{\alpha}(t_{1},t_{2}) \, .
\end{align}
The first integral gives
\begin{equation}
	C_{j}^{(\rm I)}(t) = \frac{1}{2}  \Gamma(1+\alpha) \bigl[ (1-u)^{-\alpha} - (1+u)^{-\alpha} \bigr] + \frac{1}{2} t^{\alpha} \Bigl[ \phi_{\alpha}((1-u)t) - \phi_{\alpha}((1+u)t) \Bigr] \, ,
\end{equation}
and asymptotically it reduces to
\begin{equation}
	C_{j}^{(\rm I)}(t) = \frac{1}{2}  \Gamma(1+\alpha) \bigl[ (1-u)^{-\alpha} - (1+u)^{-\alpha} \bigr] + \frac{\alpha u}{1-u^{2}}   t^{\alpha-1} + O(t^{\alpha-2}) \, .
\end{equation}
We split the second integral as already done in the previous sections writing
\begin{equation}
	C_{j}^{(\rm II)}(t) = C_{j}^{(\rm II, 1)}(t) + C_{j}^{(\rm II, 2)}(t) \, ,
\end{equation}
with
\begin{align}
	\label{} \nonumber
	C_{j}^{(\rm II,1)}(t) & =  \int_{0}^{t}\textrm{d}t_{1} \, t_{1}^{\alpha} Q_{c}^{\prime}(t-t_{1}) \int_{0}^{t} \textrm{d}t_{2} \, Q_{s}^{\prime}(t-t_{2}) \\
	& +  \int_{0}^{t}\textrm{d}t_{1} \, Q_{c}^{\prime}(t-t_{1}) \int_{0}^{t} \textrm{d}t_{2} \, t_{2}^{\alpha} Q_{s}^{\prime}(t-t_{2})
\end{align}
and
\begin{align}
	\label{} \nonumber
	C_{j}^{(\rm II,2)}(t) & = -  \int_{0}^{t}\textrm{d}t_{1} \int_{0}^{t}\textrm{d}t_{2} \, Q_{c}^{\prime}(t-t_{1}) Q_{s}^{\prime}(t-t_{2}) |t_{1}-t_{2}|^{\alpha} \, .
\end{align}
Up to exponentially small terms,

\begin{align}
	\label{} \nonumber
	C_{j}^{(\rm II, 1)}(t) & \simeq \frac{1}{2}  \bigl[ - \phi_{\alpha}((1-u)t) + \phi_{\alpha}((1+u)t) \bigr] t^{\alpha} \\
	& \simeq - \frac{\alpha u }{1-u^{2}}   t^{\alpha-1} + O(t^{\alpha-2}) \, .
\end{align}

We can now consider $C_{j}^{(\rm II,2)}(t)$, which is given by
\begin{align}
	\label{} \nonumber
	C_{j}^{(\rm II,2)}(t) & = -  \int_{0}^{t}\textrm{d}t_{1} \, \int_{0}^{t}\textrm{d}t_{2} \, Q_{c}^{\prime}(t_{1}) Q_{s}^{\prime}(t_{2}) |t_{1}-t_{2}|^{\alpha} \, .
\end{align}
In contrast with the functions $A_{j}^{(\rm II,2)}(t)$ and
$C_{j}^{(\rm II,2)}(t)$, the integrand is not symmetric under the
exchange of $t_{1}$ and $t_{2}$. This time, we approach the
calculation by performing directly the Laplace transform and then
integrating with respect to $t_{1}$ and $t_{2}$. The application of
the final value theorem then gives
\begin{equation}
	\lim_{t \rightarrow \infty} C_{j}^{(\rm II,2)}(t) = - \frac{}{4} \Gamma (1+\alpha ) \Bigl[ (1-u)^{-\alpha} - (1+u)^{-\alpha} \Bigr] \, .
\end{equation}
Assembling all the pieces together, we finally obtain

\begin{equation}
	\lim_{t \rightarrow \infty} C_{j}(t) = \frac{}{4} \Gamma (1+\alpha ) \Bigl[ (1-u)^{-\alpha} - (1+u)^{-\alpha} \Bigr] \, .
\end{equation}


\section{Characteristic functions}
\label{Appendix_Characteristic}
Here we calculate the characteristic functions $\Phi$ and $Q_{j}$
which are defined in Eqs.~(\ref{006}) and (\ref{032}), respectively.

\subsection{The function $\Phi$}
\label{appendix_phi}
By inserting Eq.~(\ref{004}) into Eq.~(\ref{006}), and recalling that
$\langle \xi_{x}(t_{1}) \xi_{y}(t_{2}) \rangle=0$, it follows that
Eq.~(\ref{006}) can be factorized as follows
\begin{equation}
\label{007}
\Phi(t, \lambda_{x},\lambda_{y}; T_{x},T_{y}; \alpha_{x}, \alpha_{y}) = \Psi(t, \lambda_{x},\lambda_{y}; T_{x} ; \alpha_{x}) \Psi(t, \lambda_{y},\lambda_{x}; T_{y} ; \alpha_{y}) \, ,
\end{equation}
where
\begin{equation}
\label{008}
\Psi(t, \lambda_{x},\lambda_{y}; T_{j} ; \alpha_{j}) = \biggl\langle \exp\biggl[ \im \int_{0}^{t} \textrm{d}\tau \, \left( \lambda_{x} Q_{c}(t-\tau) + \lambda_{y} Q_{s}(t-\tau) \right) \xi_{x}(\tau) \biggr] \biggr\rangle \, , \qquad j=x,y \, .
\end{equation}
The second factor in the r.h.s. of Eq.~(\ref{007}) is obtained upon
interchanging $\lambda_{x}$ with $\lambda_{y}$ and by using the
appropriate values for the temperature and anomalous exponent. Since
the noise is Gaussian, the characteristic function in Eq.~(\ref{008})
is completely determined by the second moment of the random process
appearing in the argument of the exponential. It is formally given by
\begin{equation}
\label{009}
G(t, \lambda_{x},\lambda_{y}; T_{j} ; \alpha_{j}) = \biggl\langle \biggl[ \im \int_{0}^{t} \textrm{d}\tau \, \left( \lambda_{x} Q_{c}(t-\tau) + \lambda_{y} Q_{s}(t-\tau) \right) \xi_{x}(\tau) \biggr]^{2} \biggr\rangle \, ,
\end{equation}
and therefore the characteristic function Eq.~(\ref{008}) is
\begin{equation}
\label{012}
\Psi(t, \lambda_{x},\lambda_{y}; T_{j} ; \alpha_{j}) = \exp\bigl[ 2^{-1}G(t, \lambda_{x},\lambda_{y}; T_{j} ; \alpha_{j})\bigr] \, .
\end{equation}
Eventually, we can write the quadratic form in $\lambda_{x}$, $\lambda_{y}$
\begin{align}
\label{010}
G(t, \lambda_{x},\lambda_{y}; T_{j} ; \alpha_{j}) & = - T_{j} \Bigl[ A_{x}(t) \lambda_{x}^{2} + B_{x}(t) \lambda_{y}^{2} + 2C_{x}(t) \lambda_{x} \lambda_{y} \Bigr] \, ,
\end{align}
By inserting Eq.~(\ref{010}) into Eq.~(\ref{012}), we have
\begin{equation}
\label{013}
\Psi(t, \lambda_{x},\lambda_{y}; T_{j} ; \alpha_{j}) = \exp\biggl[ - \frac{T_{j}}{2} \left( A_{j}(t) \lambda_{x}^{2} + B_{j}(t) \lambda_{y}^{2} + 2C_{j}(t) \lambda_{x} \lambda_{y} \right) \biggr] \, .
\end{equation}
from which Eq.~(\ref{006}) is recovered using Eq.~(\ref{007}).

\subsection{The function $\Xi_{j}$}
\label{appendix_xi}
In order to compute the characteristic function $\Xi_{j}$ it is instructive to consider a preliminary exercise which can be applied to the calculation of $\Xi_{j}$. Suppose that we want to compute the following characteristic function
\begin{equation}
\label{28112021_1614}
\Omega(\mathcal{X},t) = \langle \xi_{x}(t) \textrm{e}^{ \im (\mathcal{X} \star \xi_{x})_{t} } \rangle \, ,
\end{equation}
where $\xi_{x}(t)$ is the FGN,  $\star$ denotes the convolution defined by 
\begin{equation}
\label{28112021_1615}
(\mathcal{X} \star \xi_{x})_{t} \equiv \int_{0}^{t}\textrm{d}\tau \, \mathcal{X}(t-\tau) \xi_{x}(\tau) \, ,
\end{equation}
and $\mathcal{X}$ is some kernel. The calculation of Eq.~(\ref{28112021_1614}) is performed by expanding in power series the exponential and using the fact that only even powers of $\xi_{x}$ contributes in the averaging. Therefore
\begin{equation}
\label{28112021_1616}
\Omega(\mathcal{X},t) = \sum_{n=0}^{\infty} \frac{\im^{2n+1}}{(2n+1)!} \langle \xi_{x}(t) \langle (\mathcal{X} \star \xi_{x})_{t} )^{2n+1} \rangle \, .
\end{equation}
The quantity in angular brackets can be computed by applying Wick's theorem, which yields
\begin{equation}
\label{28112021_1620}
\langle \xi_{x}(t) \langle (\mathcal{X} \star \xi_{x})_{t} )^{2n+1} \rangle = (2n+1)!! \langle \xi_{x}(t) (\mathcal{X} \star \xi_{x})_{t} ) \rangle \Bigl[ \langle (\mathcal{X} \star \xi_{x})_{t} )^{2} \rangle \Bigr]^{n} \, ,
\end{equation}
by inserting Eq.~(\ref{28112021_1620}) into Eq.~(\ref{28112021_1616}),
\begin{equation}
\label{28112021_1623}
\Omega(\mathcal{X},t) = \im \langle \xi_{x}(t) (\mathcal{X} \star \xi_{x})_{t} ) \rangle \sum_{n=0}^{\infty} \frac{(2n+1)!!}{(2n+1)!}  \Bigl[ - \langle (\mathcal{X} \star \xi_{x})_{t} )^{2} \rangle \Bigr]^{n} 
\end{equation}
Thanks to the identity
\begin{equation}
\frac{(2n+1)!!}{(2n+1)!} = \frac{1}{2^{n}n!} \, ,
\end{equation}
it follows that Eq.~(\ref{28112021_1623}) is the Taylor expansion of the exponential. The summation of the series gives
\begin{equation}
\label{28112021_1624}
\Omega(\mathcal{X},t) = \im \langle \xi_{x}(t) (\mathcal{X} \star \xi_{x})_{t} ) \rangle \exp\biggl[ - \frac{1}{2} \langle (\mathcal{X} \star \xi_{x})_{t} )^{2} \rangle \biggr] \, .
\end{equation}

We can now to carry out the calculation of the characteristic function $\Xi_{j}$. Let us start with $\Xi_{x}$,
\begin{equation}
\Xi_{x} = \langle \xi_{x}(t) \textrm{e}^{ \im \lambda_{x}x(t) + \im \lambda_{y}y(t) } \rangle \, .
\end{equation}
The quantity in the exponential can be factorized in two terms: the first
contains $\xi_{x}$ and a second $\xi_{y}$
\begin{align}
\nonumber
\Xi_{x} & = \biggl\langle \xi_{x}(t) \exp\biggl[ \im \int_{0}^{t} \textrm{d}\tau \, \left( \lambda_{x} Q_{c}(t-\tau) + \lambda_{y} Q_{s}(t-\tau) \right) \xi_{x}(\tau) \biggr] \biggr\rangle \\
& \times \biggl\langle \exp\biggl[ \im \int_{0}^{t} \textrm{d}\tau \, \left( \lambda_{y} Q_{c}(t-\tau) + \lambda_{x} Q_{s}(t-\tau) \right) \xi_{y}(\tau) \biggr] \biggr\rangle \, .
\end{align}
The second factor is precisely
$\Psi(t, \lambda_{y},\lambda_{x}; T_{y} ; \alpha_{y})$. The first
factor has the form of Eq~.(\ref{28112021_1614}) with the kernel
$\mathcal{X}$ given by
\begin{equation}
\mathcal{X}(t-\tau) = \lambda_{x} Q_{c}(t-\tau) + \lambda_{y} Q_{s}(t-\tau) \, .
\end{equation}
The result, Eq.~(\ref{28112021_1624}), can be applied with
\begin{equation}
\langle (\mathcal{X} \star \xi_{x})_{t} )^{2} \rangle = T_{x} \left( \lambda_{x}^{2} A_{x}(t) + \lambda_{y}^{2} B_{x}(t) + 2 \lambda_{x} \lambda_{y} C_{x}(t) \right) \, ,
\end{equation}
and with
\begin{equation}
\langle \xi_{x}(t) (\mathcal{X} \star \xi_{x})_{t} ) = T_{x} \left( \lambda_{x} U_{x}(t) + \lambda_{y} V_{x}(t) \right) \, ,
\end{equation}
where $U_{j}(t)$ and $V_{j}(t)$ are the auxiliary functions
\begin{equation}
\label{034}
U_{j}(t) = \int_{0}^{t} \textrm{d}\tau \, Q_{c}(t-\tau) N_{\alpha_{j}}(t,\tau) \, , \qquad V_{j}(t) = \int_{0}^{t} \textrm{d}\tau \, Q_{s}(t-\tau) N_{\alpha_{j}}(t,\tau) \, .
\end{equation}
To recap, $\Xi_{x}$ is given by 
\begin{align}
\label{} \nonumber
\Xi_{x} & = \im T_{x} \Bigl[ \lambda_{x} U_{x}(t) + \lambda_{y} V_{x}(t) \Bigr] \exp \biggl[ - \frac{T_{x}}{2} \left( \lambda_{x}^{2} A_{x}(t) + \lambda_{y}^{2} B_{x}(t) + 2 \lambda_{x} \lambda_{y} C_{x}(t) \right) \biggr] \\
& \times \Psi(t, \lambda_{y},\lambda_{x}; T_{y}, \alpha_{y}) \, ,
\end{align}

and from Eq.~(\ref{007}) it follows that
\begin{equation}
\Xi_{x} = \im T_{x} \Bigl[ \lambda_{x} U_{x}(t) + \lambda_{y} V_{x}(t) \Bigr] \Phi(t, \lambda_{x},\lambda_{y}; T_{x},T_{y}; \alpha_{x}, \alpha_{y}) \, .
\end{equation}
An analogous reasoning applies to $\Xi_{y}$ giving
\begin{equation}
\Xi_{y} = \im T_{y} \Bigl[ \lambda_{x} V_{y}(t) + \lambda_{y} U_{y}(t) \Bigr] \Phi(t, \lambda_{x},\lambda_{y}; T_{x},T_{y}; \alpha_{x}, \alpha_{y}) \, .
\end{equation}

\subsection{The functions $U_{j}(t)$ and $V_{j}(t)$}
\label{Appendix_UV}
In order to compute the functions $U_{j}(t)$ and $V{j}(t)$ in the simplest possible way, we introduce the function
\begin{equation}
	\label{22112021_1704}
	W_{j}(t,r) = \int_{0}^{t}\textrm{d}\tau \, \textrm{e}^{-r(t-\tau)} N_{\alpha_{j}}(\tau,t) \, ,
\end{equation}
thanks to which it is possible to use the linear decomposition
\begin{align}
	\label{22112021_1948}
	U_{j}(t) & = \frac{1}{2} W_{j}(t,(1+u)) + \frac{1}{2} W_{j}(t,(1-u)) \\
	V_{j}(t) & = \frac{1}{2} W_{j}(t,(1+u)) - \frac{1}{2} W_{j}(t,(1-u)) \, .
\end{align} 

The function $W_{j}(t,r)$ has an interesting significance: it is the
equal-time correlation function between the noise of a one-dimensional
fractional Brownian motion in a fictitious optical trap. More
precisely, this identification is established by considering the
one-dimensional fractional Brownian motion in the parabolic trap such
that the system is described by the Langevin equation
\begin{equation}
	\label{22112021_1701}
	\dot{X}(t) = - r X(t) + \zeta(t) \, ,
\end{equation}
where $\zeta(t)$ is a fractional Gaussian noise with zero average and autocorrelation function given by $\langle \zeta(t_{1}) \zeta(t_{2}) \rangle = N_{\alpha_{j}}(t_{1},t_{2})$, the parameter $r$ is the stiffness of the potential. In free space, $N_{\alpha_{j}}(t_{1},t_{2})$ originates the autocorrelation function of the position
\begin{equation}
	\langle X_{f}(t_{1}) X_{f}(t_{2}) \rangle =  \left( t_{1}^{\alpha} + t_{2}^{\alpha} - |t_{1}-t_{2}|^{\alpha} \right) \, .
\end{equation}
The formal solution of Eq.~(\ref{22112021_1701}) with initial condition $X(0)=0$ is
\begin{equation}
	\label{22112021_1702}
	X(t) = \int_{0}^{t}\textrm{d}\tau \, \textrm{e}^{-r(t-\tau)} \zeta(\tau) \, .
\end{equation}
We multiply both sides of Eq.~(\ref{22112021_1702}) by the noise and then we perform the average; the result is
\begin{equation}
	\label{22112021_1703}
	\langle X(t) \zeta(t) \rangle = \int_{0}^{t}\textrm{d}\tau \, \textrm{e}^{-r(t-\tau)} \langle \zeta(\tau) \zeta(t) \rangle\, .
\end{equation}
It is now clear that the right hand side of Eq.~(\ref{22112021_1703}) is the function $W_{j}(t,r)$. As a result, we can write
\begin{equation}
	\label{22112021_1704_v2}
	W_{j}(t,r) = \langle X(t) \zeta(t) \rangle \, ,
\end{equation}
as we anticipated. We then plug in Eq.~(\ref{22112021_1704_v2}) the
expression for $\zeta(t)$ provided by Eq.~(\ref{22112021_1701}) to get
\begin{equation}
	\label{22112021_1705}
	W_{j}(t,r) = r \langle X^{2}(t) \rangle + \langle X(t) \dot{X}(t) \rangle \, .
\end{equation}
Now the problem is reduced to the evaluation of the mean square displacement and the position-velocity correlation function $\langle X(t) \dot{X}(t) \rangle$. The latter can be computed directly from the correlation function as follows
\begin{equation}
	\label{22112021_1706}
	\langle X(t) \dot{X}(t) \rangle = \partial_{t^{\prime}} \langle X(t) X(t^{\prime}) \rangle \big\vert_{t^{\prime}=t} \, .
\end{equation}

It follows that the MSD is
\begin{equation}
	\langle X^{2}(t) \rangle = 2  \textrm{e}^{- t} t^{\alpha } {}_{1}F_{2}\left(1,\frac{1+\alpha}{2},1+\frac{\alpha}{2};t^{2}/4\right) \,
\end{equation}
and
\begin{equation}
	\lim_{t \rightarrow \infty} \langle X^{2}(t) \rangle =  \Gamma(1+\alpha) \,.
\end{equation}
For the equal-time velocity-position correlation function we have
	\begin{align}
	\label{} \nonumber
	\langle X(t) \dot{X}(t) \rangle & =  t^{\alpha-1} \textrm{e}^{-t} \biggl[ -t \, {}_{1}F_{2}\left(1,\frac{1+\alpha}{2},1+\frac{\alpha}{2};t^{2}/4\right) + \alpha \, {}_{1}F_{2}\left(1,\frac{1+\alpha}{2},\frac{\alpha}{2};t^{2}/4\right) \biggr]
\end{align}
and
\begin{equation}
	\lim_{t \rightarrow \infty} \langle X(t) \dot{X}(t) \rangle = 0 \, .
\end{equation}
In summary,
\begin{equation}
	W_{j}(t\rightarrow \infty,r) =  \Gamma(1+\alpha_{j}) \, .
\end{equation}
Coming back to Eq.~(\ref{22112021_1948}), the infinite-time limit gives
\begin{align}
	\label{22112021_1949}
	\lim_{t \rightarrow \infty} U_{j}(t) & =  \Gamma(1+\alpha_{j}) \\
	\lim_{t \rightarrow \infty} V_{j}(t) & = 0 \, .
\end{align} 

\section{Calculation of the current vector}
\label{appendix_Current}
Let us introduce the following notation
\begin{equation}
\llbracket f(\lambda_{x},\lambda_{y}) \rrbracket = \int_{\mathbb{R}^{2}}\frac{\textrm{d}\lambda_{x}\textrm{d}\lambda_{y}}{(2\pi)^{2}} \textrm{e}^{ - \im x \lambda_{x} - \im y \lambda_{y} } f(\lambda_{x},\lambda_{y}) \, .
\end{equation}
Because the characteristic function $\Phi(t, \lambda_{x},\lambda_{y})$ is a bivariate Gaussian distribution, it is easy to derive the following results
\begin{align}
\llbracket \Phi(t, \lambda_{x},\lambda_{y}) \rrbracket & = P(x,y,t) \\
\llbracket \im \partial_{\lambda_{x}} \Phi(t, \lambda_{x},\lambda_{y}) \rrbracket & = - x P(x,y,t) \\
\llbracket \im \partial_{\lambda_{y}} \Phi(t, \lambda_{x},\lambda_{y}) \rrbracket & = - y P(x,y,t) \\
\llbracket \im \lambda_{x}\Phi(t, \lambda_{x},\lambda_{y}) \rrbracket & = \frac{b(t)x-c(t)y}{a(t)b(t)-c^{2}(t)} P(x,y,t) \\
\llbracket \im \lambda_{y} \Phi(t, \lambda_{x},\lambda_{y}) \rrbracket & = \frac{-c(t)x+a(t)y}{a(t)b(t)-c^{2}(t)} P(x,y,t) \, .
\end{align}
Using these results, one obtains 
\begin{align}
	\label{}
	J_{x}(x, y; t) & = \biggl[ - x + u y + T_{x}U_{x}(t) \frac{b(t)x-c(t)y}{a(t)b(t)-c^{2}(t)} + T_{x}V_{x}(t) \frac{-c(t)x+a(t)y}{a(t)b(t)-c^{2}(t)}  \biggr] P(x,y,t) \\
	J_{y}(x, y; t) & = \biggl[ - y + u x + T_{y}V_{y}(t) \frac{b(t)x-c(t)y}{a(t)b(t)-c^{2}(t)} + T_{y}U_{y}(t) \frac{-c(t)x+a(t)y}{a(t)b(t)-c^{2}(t)}  \biggr] P(x,y,t) \, .
\end{align}

In the infinite-time limit the function $V_{j}$ vanishes and $U_j(t) \rightarrow \Gamma(1+\alpha_{j})$ and we are left with
\begin{align}
	\label{}
	J_{x}(x, y) & = \biggl[ - x + u y +  \Gamma(1+\alpha_{x}) T_{x} \frac{bx-cy}{ab-c^{2}} \biggr] P(x,y) \\
	J_{y}(x, y) & = \biggl[ - y + u x -  \Gamma(1+\alpha_{y}) T_{y}\frac{cx-ay}{ab-c^{2}}  \biggr] P(x,y) \, .
\end{align}
 We observe that the quantities in square brackets are related to the potential energy $U(x,y)$ and to the density $P(x,y)$ via
\begin{equation}
	\partial_{x}U(x,y) = x - uy \, , \qquad \partial_{y}U(x,y) = y - ux \, ,
\end{equation}
and
\begin{equation}
	\frac{\partial_{x}P(x,y)}{P(x,y)} = - \frac{bx-cy}{ab-c^{2}} \, , \qquad \frac{\partial_{y}P(x,y)}{P(x,y)} = \frac{cx-ay}{ab-c^{2}} \, ,
\end{equation}
which means that we can write the currents as Eq.~(\ref{eq:currents}).


\section*{References}


\providecommand{\newblock}{}

\end{document}